\documentclass[12pt]{iopart}

\usepackage{iopams}  
\usepackage{graphicx} 
\graphicspath{{./}} 
\usepackage{booktabs} 
\usepackage[font=small,labelfont=bf]{caption} 
\usepackage{amsfonts, amsthm, amssymb} 
\usepackage{subcaption}
\usepackage{dcolumn}
\usepackage{xcolor}
\usepackage{hyperref}
\usepackage[normalem]{ulem}

\begin{document}

\title[Spritz: a new fully general-relativistic MHD code]{Spritz: a new fully general-relativistic magnetohydrodynamic code}

\author{F. Cipolletta$^{1, \, 2, \, 3}$,  J. V. Kalinani$^{4, \, 5}$, B. Giacomazzo$^{6, \, 7, \, 8 \, \ast}$, \\R. Ciolfi$^{9, \, 5}$}

\address{
$^{1}$INFN-TIFPA, Trento Institute for Fundamental Physics and Applications, Via Sommarive 14, I-38123 Trento, Italy
\\$^{2}$Dipartimento di Fisica, Universit\`a di Trento, Via Sommarive 14, I-38123 Trento, Italy
\\$^{3}$Center for Computational Relativity and Gravitation, School of Mathematical Sciences, Rochester Institute of Technology, 85 Lomb Memorial Drive, Rochester, New York 14623, USA
\\$^{4}$Universit\`a di Padova, Dipartimento di Fisica e Astronomia, Via Francesco Marzolo 8, I-35131 Padova, Italy
\\$^{5}$INFN, Sezione di Padova, Via Francesco Marzolo 8, I-35131 Padova, Italy
\\$^{6}$Dipartimento di Fisica G. Occhialini, Universit\`a di Milano - Bicocca, Piazza della Scienza 3, I-20126 Milano, Italy
\\$^{7}$INFN, Sezione di Milano Bicocca, Piazza della Scienza 3, I-20126 Milano, Italy
\\$^{8}$INAF, Osservatorio Astronomico di Brera, Via E. Bianchi 46, I-23807 Merate, Italy
\\$^{9}$INAF, Osservatorio Astronomico di Padova, Vicolo dell'Osservatorio 5, I-35122 Padova, Italy
}

\ead{$^{\ast}$\texttt{bruno.giacomazzo@unimib.it}}
\vspace{10pt}
\begin{indented}
\item[]December 2019
\end{indented}

\begin{abstract}
The new era of multimessenger astrophysics requires the capability of studying different aspects of the evolution of compact objects. In particular, the merger of neutron star binaries is a strong source of gravitational waves and electromagnetic radiation, from radio to $\gamma$-rays, as demonstrated by the detection of GW170817 and its electromagnetic counterparts. In order to understand the physical mechanisms involved in such systems, it is necessary to employ fully general relativistic magnetohydrodynamic (GRMHD) simulations able to include the effects of a composition and temperature dependent equation of state describing neutron star matter as well as neutrino emission and reabsorption.
Here, we present our new code named \texttt{Spritz} that solves the GRMHD equations in 3D Cartesian coordinates and on a dynamical spacetime. 
The code can support tabulated equations of state, taking into account finite temperature effects and allowing for the inclusion of neutrino radiation. In this first paper, we present the general features of the code and a series of tests performed in special and general relativity to assess the robustness of the basic GRMHD algorithms implemented. Among these tests, we also present the first comparison between a non-staggered and a staggered formulation of the vector potential evolution, which is used to guarantee the divergence-less character of the magnetic field. With respect to other publicly available GRMHD codes, \texttt{Spritz} combines the robust approach of a staggered formulation of the vector potential together with the use of an equation of state driver (\texttt{EOS\_Omni}) that can allow the code to use finite temperature equations of state. A next version of the code will fully test the \texttt{EOS\_Omni} driver by coupling it with a neutrino leakage scheme.
\end{abstract}

%
%
\submitto{\CQG}
%
%
%
\section{Introduction}
\label{Intro}
Magnetic fields play a crucial role in several high-energy
astrophysical scenarios at different scales, from Active Galactic
Nuclei (AGN) to Gamma-Ray Bursts (GRBs). 
These
phenomena involve compact objects such as neutron stars (NSs) and black holes (BHs) and therefore any attempt to model them requires a general relativistic treatment.
As a consequence, studying this kind of systems demands to solve the full set of 
general relativistic magnetohydrodynamic (GRMHD) equations~\cite{anton2006numerical}.
In most situations, the GRMHD equations are to be solved numerically, 
often on dynamical spacetimes, and a number of GRMHD codes have 
been developed over the years for this purpose (e.g., ~\cite{giacomazzo2007whiskymhd, mosta2013grhydro, etienne2015illinoisgrmhd}). 
Some of them have been used, in particular, to study compact binary mergers 
(e.g., \cite{Liu2008, Anderson2008, giacomazzo2011accurate, Kiuchi2018, Ciolfi2019}) and accretion onto supermassive BHs 
(e.g., \cite{Palenzuela2010, Giacomazzo2012, Farris2012, Gold2014, Kelly2017, dascoli2018, EHTCodeComparison}). 

In the case of compact binary mergers, GRMHD codes have been used to
simulate NS-NS and NS-BH mergers in order to study the effects of
magnetic fields on the gravitational wave (GW) and electromagnetic
(EM) emission (e.g., \cite{Kawamura2016, Ciolfi2017}). For instance, GRMHD simulations have recently provided indications that, under certain conditions, the BH remnant of a NS-NS or NS-BH merger may be able to give
rise to a relativistic jet and hence power a short
GRB~\cite{Paschalidis2015, Ruiz2016}. 
This is a likely scenario to explain the connection between compact binary mergers and short 
GRBs, recently confirmed by the first simultaneous observation 
of GWs emitted by a NS-NS merger and a
short GRB~\cite{PhysRevLett.119.161101,LVC-GRB}.
Concerning the accretion onto supermassive BH mergers, current simulations aim at predicting the light curves of possible EM counterparts of future GW sources detected by LISA~\cite{Schnittman2013, Amaro-Seoane2017}.

In this paper, we present our new fully GRMHD numerical code, named
\texttt{Spritz}, that solves the GRMHD equations in 3D and on a
dynamical spacetime. The code inherits a number of basic features from the
\texttt{WhiskyMHD} code~\cite{giacomazzo2007whiskymhd}, but it also takes
advantage of methods implemented and tested in the publicly available
\texttt{GRHydro}~\cite{mosta2013grhydro} and
\texttt{IllinoisGRMHD}~\cite{etienne2015illinoisgrmhd} codes.
The \texttt{WhiskyMHD} code has been used successfully to simulate
NS-NS mergers~\cite{giacomazzo2011accurate, Ciolfi2019, Kawamura2016, Ciolfi2017, Giacomazzo2009, Rezzolla2011, GiacomazzoPerna2013, Giacomazzo2015, Endrizzi2016} and accretion onto supermassive black hole binaries~\cite{Giacomazzo2012BBH}, but it is limited to the use of simple
piecewise polytropic equations of state~\cite{Read2009} and it is not
able to take into account neutrino emission. 
Moreover, this code can evolve the
vector potential instead of the magnetic field, but employing a non-staggered formalism  that may have undesired effects on the evolution (see discussion in the following sections). 

The new \texttt{Spritz} code can instead handle any equation of state
for which the pressure is a function of rest-mass density,
temperature, and electron fraction and therefore can also use modern
tabulated equations of state. This has been possible by following a
similar approach used in the \texttt{GRHydro} code, which can use
finite temperature tabulated equations of state, but that still lacks
of a magnetic field implementation able to handle correctly the use of
mesh refinement techniques. \texttt{Spritz} also implements a
staggered version of the vector potential formulation in a formalism
that, as discussed later in the paper, recovers the original
conservative flux-CT approach implemented in the original version of
\texttt{WhiskyMHD}. This has been possible by using algorithms similar
to those implemented in \texttt{IllinoisGRMHD}, which at the moment
can only handle simple ideal fluid equations of state.
Therefore, the \texttt{Spritz} code aims at merging together the main   
capabilities of the three codes mentioned above. The use in particular of a new equation of state driver will allow \texttt{Spritz} to implement neutrino radiation via a leakage scheme currently under testing and that will be presented in a future paper.

Here, we present a series of extensive
tests in 1D, 2D, and 3D, including, for the first time, a comparison between 
staggered and non-staggered schemes for the vector potential evolution and a
rather demanding spherical explosion test. The \texttt{Spritz} code passes
all the tests successfully and it will be soon used to carry out NS-NS
and NS-BH merger simulations. The current version of the code is publicly
available and it can be downloaded from Zenodo~\cite{zenodo-spritz}.

The paper is organized as follows: in \Sref{sec2} we present the GRMHD
equations and the formulation used in the code; in \Sref{sec3} the
main numerical methods are discussed; in \Sref{sec4} we present the
results of our tests; and in \Sref{sec5} we summarize the main results
and discuss future developments. We use a system of units such that
$G=c=1$ unless otherwise specified. Greek indices run from 0 to 3 and
Latin indices run from 1 to 3.

\section{Equations}
\label{sec2}

In this section we summarize the theoretical background and the equations implemented in \texttt{Spritz}, giving also the main references for the reader who wants to go deeper in the related details. 
In addition to these references, it is worth to mention the book \cite{baumgarte2010numerical} which presents an extensive theoretical introduction to numerical relativity approaches to solving Einstein's Equations in several physical scenarios.

\subsection{3+1 spacetime formulation}
\label{3+1}

Our numerical methods and implementation are largely based on the ones employed in \texttt{WhiskyMHD} \cite{giacomazzo2007whiskymhd}, where a 3+1 formulation of the Einstein's equations is adopted. In such a framework, the form of the line element is:
\begin{equation}
\label{LinEl}
ds^{2} = g_{\mu \nu} dx^{\mu} dx^{\nu} = -\left( \alpha^{2} - \beta^{i}\beta_{i} \right) dt^2 + 2 \beta_{i} dx^{i} dt + \gamma_{ij} dx^{i}dx^{j},
\end{equation}
where the usual Einstein notation is adopted. Here $g_{\mu \nu}$ is the metric tensor, $\gamma_{ij}$ its purely spatial part, and $\alpha$ and $\beta^{i}$ are respectively the \textit{lapse} and the \textit{shift} vector. Our coordinate setting considers $x^{0} \equiv t$.

\texttt{Spritz} makes use of the conservative formulation presented in \cite{anton2006numerical}, which is the GRMHD version of the original general relativistic hydrodynamics Valencia formulation~\cite{banyuls1997numerical,marti1991numerical}. Here, the natural observer is called the \textit{Eulerian observer} and its four--velocity $\bi{n}$ is normal to the 3--dimensional hypersurface of constant $t$ with the following components:
\begin{equation}
\label{natvel}
\eqalign{
n^{\mu} &= \frac{1}{\alpha} \left( 1, -\beta^{i} \right), \\
n_{\mu} &= \left( -\alpha,0,0,0 \right).
}
\end{equation}

When considering matter, the spatial components of the fluid velocity measured by the Eulerian observer read
\begin{equation}
\label{spatfluidvel}
v^{i} = \frac{h^{i}_{\mu} u^{\mu}}{-\bi{u} \cdot \bi{n}} = \frac{u^i}{\alpha u^t} + \frac{\beta^i}{\alpha} = \frac{u^i}{W} + \frac{\beta^i}{\alpha},
\end{equation}
where $\bi{u}$ is the fluid four--velocity, $h_{\mu \nu} = g_{\mu \nu} + n_{\mu} n_{\nu}$ is the projector onto the aforementioned hypersurface at constant $t$, $W = 1/\sqrt{1-v^2}$ is the Lorentz factor, and $v^2\equiv\gamma_{ij} v^i v^j$ is the square norm of $\bi{v}$.

\subsection{Electromagnetic field}
\label{sec:Maxwell}

The general relativistic formulation of \cite{anton2006numerical} describes the electromagnetic field via the Faraday tensor and its dual, given respectively by
\begin{equation}
\label{Faraday}
F^{\mu \nu} = U^{\mu} E^{\nu} - U^{\nu} E^{\mu} - \eta^{\mu \nu \lambda \delta} U_{\lambda} B_{\delta},
\end{equation}
\begin{equation}
\label{FaradayDual}
^{*}F^{\mu \nu} = \frac{1}{2} \eta^{\mu \nu \lambda \delta} F_{\lambda \delta} = U^{\mu} B^{\nu} - U^{\nu} B^{\mu} - \eta^{\mu \nu \lambda \delta} U_{\lambda} E_{\delta},
\end{equation}
being $E^{\mu}$ the electric field, $B^{\mu}$ the magnetic field, $U^{\mu}$ a generic observer's four--velocity, and $\eta^{\mu \nu \lambda \delta} = \frac{1}{\sqrt{-g}} \left[ \mu \nu \lambda \delta \right]$ the volume element.

The equations governing the electromagnetic field and its evolution are the well known Maxwell's equations
\begin{equation}
\label{eq:Maxwell}
\eqalign{
{\nabla}_{\nu} ^{*}F^{\mu \nu} &= 0  \, , \\
{\nabla}_{\nu} F^{\mu \nu} &= 4 \pi \mathcal{J}^{\mu} \, ,
}
\end{equation}
where $\bi{\mathcal{J}}$ is the four--vector current density, which can be expressed through the Ohm's law as
\begin{equation}
\label{4curr}
\mathcal{J}^{\mu} = q u^{\mu} + \sigma F^{\mu \nu} u_{\nu} \, ,
\end{equation}
with $q$ the proper charge density and $\sigma$ the electric conductivity. In the ideal MHD regime (i.e., when $\sigma \to \infty$ and $F^{\mu \nu} u_{\nu} \to 0$) \Eref{Faraday} and \Eref{FaradayDual} can be expressed as
\begin{equation}
\label{FaradayIdeal}
F^{\mu \nu} = \eta^{\alpha \beta \mu \nu} b_{\alpha} u_{\beta}, \qquad
^{*}F^{\mu \nu} = b^{\mu} u^{\nu} - b^{\nu} u^{\mu} = \frac{u^{\mu} B^{\nu} - u^{\nu} B^{\mu}}{W} \, , 
\end{equation}
where $\bi{b}$ is the magnetic field according to the comoving observer, which can be written component--wise as follows~\cite{giacomazzo2007whiskymhd}:
\begin{equation}
\label{bsmall}
b^0 = \frac{W B^i v_i}{\alpha}, \qquad b^i = \frac{B^i + \alpha b^0 u^i}{W}, \qquad b^2 \equiv b^\mu b_\mu = \frac{B^2 + \alpha^2 \left( b^0 \right)^2}{W^2} \, .
\end{equation}
Here, $B^2 \equiv B^i B_i$, where $\bi{B}$ is now the magnetic field measured by the Eulerian observer (i.e., from now on $U^\mu=n^\mu$). With \Eref{FaradayIdeal}, the Maxwell's equations considering the dual of Faraday tensor can be written as
\begin{equation}
\label{MaxwellIdeal}
{\nabla}_{\nu} ^{*}F^{\mu \nu} = \frac{1}{\sqrt{-g}} \partial_{\nu} \left( \sqrt{-g} \left(b^{\mu} u^{\nu} - b^{\nu} u^{\mu} \right) \right) = 0.
\end{equation}

Splitted in its different components, \Eref{MaxwellIdeal} provides the equations governing the magnetic field constraints and evolution, namely the \textit{divergence--free condition}
\begin{equation}
\label{divfree}
\partial_i \tilde{B}^i = 0 \, ,
\end{equation}
where $\tilde{B}^i \equiv \sqrt{\gamma} B^i$, and the magnetic field \textit{induction equations}
\begin{equation}
\label{magnind}
\partial_t \tilde{B}^i = \partial_j \left( \tilde{v}^i \tilde{B}^j - \tilde{v}^j \tilde{B}^i \right) \, ,
\end{equation}
where $\tilde{v}^i \equiv \alpha v^i - \beta ^i$.

\subsection{Conservative approach}
\label{Cons}

The stress--energy tensor, considering a perfect fluid and the contribution of the magnetic field, can be written as
\begin{equation}
\label{Tmunu}
T^{\mu \nu} = \left( \rho h + b^2 \right) u^{\mu} u^{\nu} + \left( p_\mathrm{gas} + p_\mathrm{mag} \right) g^{\mu \nu} - b^{\mu} b^{\nu},
\end{equation}
being $\rho$ the rest--mass density, $p_\mathrm{gas}$ the gas pressure, $p_\mathrm{mag} \equiv \frac{b^2}{2}$ the magnetic pressure, $h = 1 + \varepsilon + \frac{p_\mathrm{gas}}{\rho}$ the relativistic specific enthalpy, and $\varepsilon$ the specific internal energy.

The energy-momentum conservation
\begin{equation}
\label{consTmunu}
\nabla_{\nu} T^{\mu \nu} = 0 \, ,
\end{equation}
the conservation of baryon number
\begin{equation}
\label{consbaryon}
\nabla_{\nu} \left( \rho u^{\nu} \right) = 0 \, ,
\end{equation}
Maxwell's equations for the magnetic field~(\ref{magnind}), and an equation of state (EOS, see~\ref{EOS}) give together  the complete set of equations describing the evolution of the primitive variables, i.e., $\bi{U} = \left[ \rho, v_j, \varepsilon, B^k \right]$. As usual, these equations can be written in the following conservative form:
\begin{equation}
\label{consequat}
\frac{1}{\sqrt{-g}} \left[ \partial_t \left( \sqrt{\gamma} \bi{F}^0 \right) + \partial_i \left( \sqrt{-g} \bi{F}^i \right) \right] = \bi{S},
\end{equation}
being $\bi{F}^0 \equiv \left[ D, S_j, \tau, B^k \right]$ the vector of conserved variables, defined in terms of the primitive ones as
\begin{equation}
\label{P2Csystem}
\eqalign{
D &\equiv \rho W, \\
S_{j} &\equiv \left( \rho h + b^2 \right) W^2 v_{j} - \alpha b^0 b_{j}, \\
\tau &\equiv \left( \rho h + b^2 \right) W^2 - \left( p_\mathrm{gas} + p_\mathrm{mag} \right) - \alpha^2 \left( b^0 \right)^2 - D, \\
B^k &\equiv B^k,
}
\end{equation}
$\bi{F}^i$ the vector of fluxes defined as
\begin{equation}
\label{conservedvector}
\bi{F}^i \equiv \left[ \eqalign{
&\qquad D\tilde{v}^i / \alpha \\
S_j \tilde{v}^i / \alpha + &\left( p_\mathrm{gas} + p_\mathrm{mag} \right) \delta^i_j - b_j B^i / W \\
\tau \tilde{v}^i / \alpha + &\left( p_\mathrm{gas} + p_\mathrm{mag} \right) v^i - \alpha b^0 B^i / W \\
&B^k \tilde{v}^i / \alpha - B^i \tilde{v}^k / \alpha
} \right]\,,
\end{equation}
and $\bi{S}$ the vector of sources that reads
\begin{equation}
\label{sources}
\bi{S} \equiv \left[ \eqalign{
& \qquad 0 \\
T^{\mu \nu} &\left( \partial_{\mu} g_{\nu j} - \Gamma^{\delta}_{\nu \mu} g_{\delta j} \right) \\
\alpha &\left( T^{\mu 0} \partial_{\mu} \ln{\alpha} - T^{\mu \nu} \Gamma^{0}_{\nu \mu} \right) \\
& \qquad 0^k
} \right].
\end{equation}
In order to avoid time derivatives of the metric in the source terms, these are rewritten as done in the case of the \texttt{Whisky} code~\cite{baiotti2005three} (see section 4.3.2 of ~\cite{BaiottiPhDThesis} for details).

\subsection{Electromagnetic gauge conditions}
\label{GaugeCond}

In order to accurately describe the magnetic field and its evolution, it can be convenient to formulate the problem in terms of the vector potential (see, e.g., \cite{feynman1979feynman}). Considering $\nabla$ as a purely spatial operator, one may write
\begin{equation}
\label{rotB}
\bi{B} = \nabla \times \bi{A} \, ,
\end{equation}
so that
\begin{equation}
\label{inductsatisfy}
\nabla \cdot \bi{B} = \nabla \cdot \left( \nabla \times \bi{A} \right) = 0 \, ,
\end{equation}
and thus evolving the vector potential $\bi{A}$ will automatically satisfy \Eref{divfree}.

As already done in \cite{baumgarte2003collapse,baumgarte2003general,etienne2012relativistic}, we then introduce the four--vector potential
\begin{equation}
\label{4vector}
\mathcal{A}_{\nu} = n_{\nu} \Phi + A_{\nu} \, ,
\end{equation}
being $A_{\nu}$ the purely spatial vector potential and $\Phi$ the scalar potential. With this, \Eref{divfree} and \Eref{magnind} become respectively
\begin{equation}
\label{Adivfree}
B^i = \epsilon^{ijk} \partial_j A_k \, ,
\end{equation}
and
\begin{equation}
\label{Ainduct}
\partial_t A_i = -E_i - \partial_i \left( \alpha \Phi - \beta^j A_j \right),
\end{equation}
where $\epsilon^{ijk} = n_{\nu} \epsilon^{\nu ijk}$ is the three--dimensional spatial Levi--Civita tensor.

However, the choice of the four-vector potential $\mathcal{A}^{\nu}$ is not unique and one has to choose a specific gauge. The first GRMHD simulations that employed the vector potential as an evolution variable were performed using the \textit{algebraic} gauge~\cite{etienne2012relativistic,etienne2010relativistic} where the scalar potential satisfy the following equation:
\begin{equation}
\label{algebraic}
\Phi = \frac{1}{\alpha} \left( \beta^j A_j \right).
\end{equation}
In this way~\Eref{Ainduct} is strongly simplified, being reduced to
\begin{equation}
\label{Ainduct-algebraic}
\partial_t A_i = - E_i \, ,
\end{equation}
and therefore it does not require to evolve the scalar potential $\Phi$.

More recently, GRMHD simulations started to use the \textit{Lorenz} gauge~\cite{etienne2012relativistic}, which consists of imposing the constraint $\nabla_{\nu} \mathcal{A}^{\nu} = 0$. This gauge requires now to solve also the evolution equation for the scalar potential:
\begin{equation}
\label{lorenz}
\partial_t \left( \sqrt{\gamma} \Phi \right) + \partial_i \left( \alpha \sqrt{\gamma} A^i - \sqrt{\gamma} \beta^i \Phi \right) = 0.
\end{equation}

The \textit{Lorenz} gauge has been shown to perform better in those simulations that implement adaptive mesh refinement, such as, for example, binary neutron star and neutron star--black hole mergers~\cite{etienne2012relativistic}. 
The \textit{algebraic} gauge may indeed cause interpolation errors at the boundaries between refinement levels and thus produce spurious magnetic fields (see~\cite{etienne2012relativistic} for more details). An even more robust gauge choice has been introduced in~\cite{Farris2012} with the name of \textit{generalized Lorenz} gauge:
\begin{equation}
  \nabla_{\nu} \mathcal{A}^{\nu} = \xi n_\nu \mathcal{A}^{\nu} \, ,
\end{equation}
where $\xi$ is a parameter that is typically set to be equal to $1.5/\Delta t_{\rm max}$, being $\Delta t_{\rm max}$ the timestep of the coarsest refinement level~\cite{etienne2015illinoisgrmhd}. When employing this gauge choice the evolution equation for the scalar potential becomes
\begin{equation}
\label{lorenz-generalized}
\partial_t \left( \sqrt{\gamma} \Phi \right) + \partial_i \left( \alpha \sqrt{\gamma} A^i - \sqrt{\gamma} \beta^i \Phi \right) = -\xi \alpha \sqrt{\gamma} \Phi \, .
\end{equation}
In \texttt{Spritz} we adopt the \textit{generalized Lorenz} gauge which is also the gauge used in the latest \texttt{WhiskyMHD} simulations~\cite{Ciolfi2019, Kawamura2016, Ciolfi2017, GiacomazzoPerna2013, Giacomazzo2015, Endrizzi2016}.

\section{Numerical Implementation}
\label{sec3}

In the present section we summarize the main numerical methods implemented within the \texttt{Spritz} code. The code is based on the Einstein Toolkit~\cite{ETKpaper, EinsteinToolkit:2019_10} which provides a framework to automatically parallelize the code for the use on supercomputers as well as a number of open-source codes providing a number of useful routines, such as those for the evolution of the spacetime, adaptive mesh refinement, input and output of data, checkpointing, and many others. 

\subsection{Riemann Solvers}
\label{RieHRSC}

The \texttt{Spritz} code adopts High Resolution Shock Capturing (HRSC) methods to solve \Eref{consequat}. These methods are based on the choice of reconstruction algorithms, to compute the values of primitive variables at the interface between numerical cells, and of approximate Riemann solvers, to finally compute the fluxes.

Our default Riemann solver is the Harten--Lax--van--Leer--Einfeldt (HLLE) \cite{harten1983upstream}, where the numerical fluxes at cell interfaces are computed as follows:
\begin{equation}
\label{HLLEfluxes}
\bi{F}^i = \frac{c_\mathrm{min} \bi{F}^i_\mathrm{r} + c_\mathrm{max} \bi{F}^i_\mathrm{l} - c_\mathrm{max}  c_\mathrm{min} \left( \bi{F}^0_\mathrm{r} - \bi{F}^0_\mathrm{l} \right)}{c_\mathrm{max} + c_\mathrm{min}} \, ,
\end{equation}
where a subscript r (l) means that the function is computed at the right (left) side of the cell interface and 
$c_\mathrm{max} \equiv \max \left( 0, c_{+,\tiny{\mbox{l}}}, c_{+,\tiny{\mbox{r}}} \right)$,
$c_\mathrm{min} \equiv -\min \left( 0, c_{-,\tiny{\mbox{l}}}, c_{-,\tiny{\mbox{r}}} \right)$, where 
$c_{\pm,\tiny{\mbox{r}}}$ ($c_{\pm,\tiny{\mbox{l}}}$) are the right-going ($+$) and left-going ($-$) maximum wave speeds computed from the primitive variables $\bi{U}$ at the right (left) side.

We decided also to implement the Lax--Friedrichs (LxF) scheme \cite{toro2013riemann}, that is
\begin{equation}
\label{LxFfluxes}
\bi{F}^i = \frac{\bi{F}^i_\mathrm{r} + \bi{F}^i_\mathrm{l} - c_\mathrm{wave} \left( \bi{F}^0_\mathrm{r} - \bi{F}^0_\mathrm{l} \right)}{2},
\end{equation}
where $c_\mathrm{wave} = \max(c_\mathrm{max}, c_\mathrm{min})$~\cite{del2003efficient}. This scheme is a very dissipative one and it can be useful in cases where strong jumps in pressure must be considered.

In order to compute the values of $\bi{F}^0$ at right and left sides of cell's interfaces from the primitive variables, we adopt the third--order Piece-wise Parabolic Method (\textit{PPM}) \cite{colella1984piecewise}. In addition, for those cases that require more dissipative methods, for example in presence of strong shocks, we also implemented the second--order total variation diminishing (TVD) \textit{minmod} method \cite{toro2013riemann}.

\subsection{Electromagnetic Field Evolution}
\label{Stagg}

As already mentioned in \Sref{GaugeCond}, the \texttt{Spritz} code is meant to deal with different electromagnetic gauge conditions for the vector potential.

In order to accurately evolve the magnetic field, particular care must be taken in solving numerically \Eref{Ainduct-algebraic}, in the case of the algebraic gauge, or \Eref{Ainduct} and \Eref{lorenz-generalized}, in case of the generalized Lorenz gauge. From now on, we will also use the following definition for simplicity:
\begin{equation}
\label{psimhd}
\Psi_\mathrm{mhd} \equiv \sqrt{\gamma} \Phi \,.
\end{equation}

\begin{figure}[b!]
\begin{center} 
\includegraphics[width=0.5\linewidth]{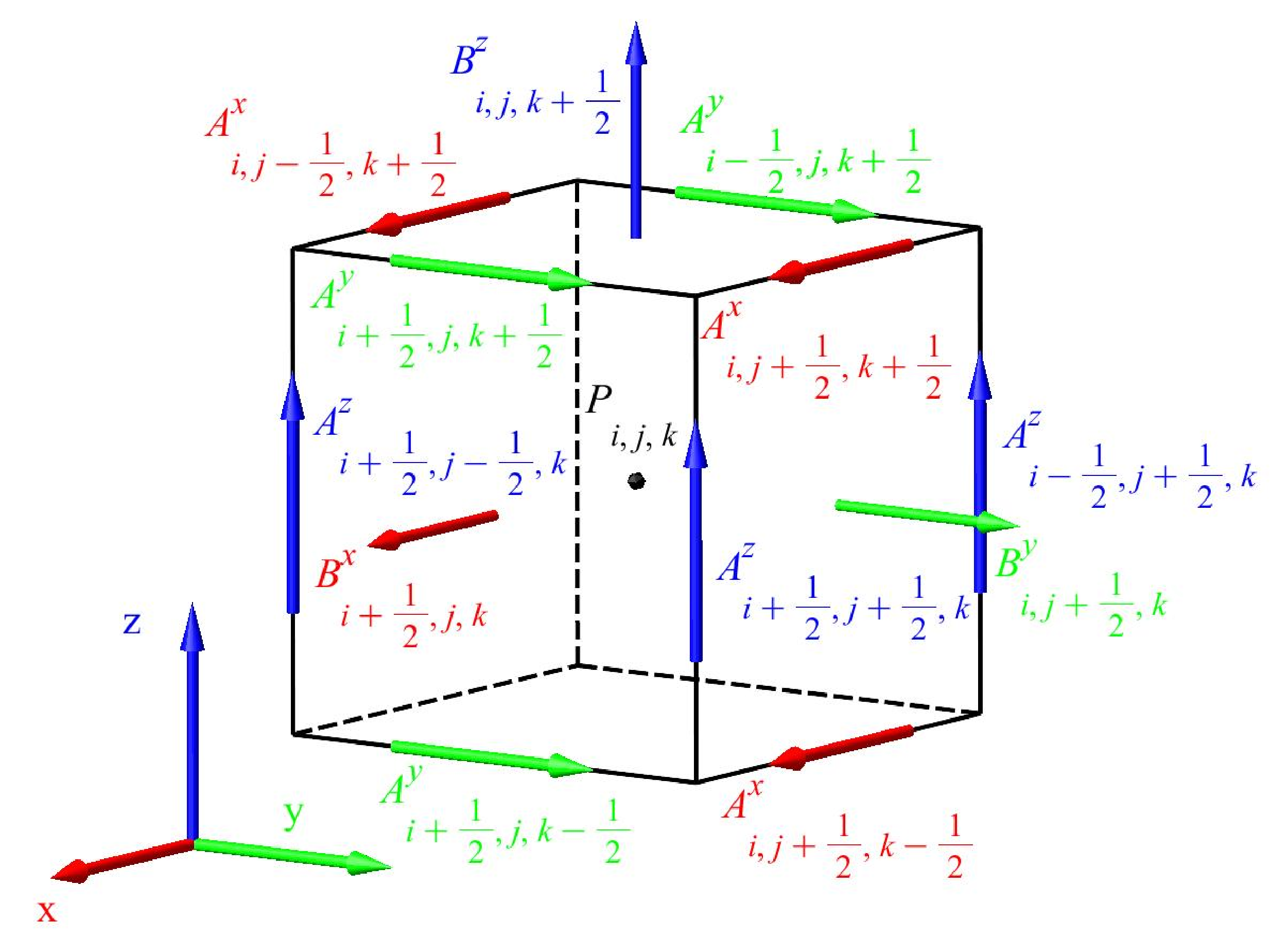}
\captionof{figure}{\label{Fig1}Representation of storage locations for magnetic field and vector potential components in our numerical code. Point $P_{i,j,k}$ represents the cell's center.}
\end{center}
\end{figure}

As in every numerical code, the spatial domain is divided in grid--cells of user specified dimensions. The fluid's state variables (e.g., $\rho$, $p_{gas}$, $\bi{v}$) are stored in the grid--cell's centers. Usually, the electric and magnetic fields ($\bi{E}$ and $\bi{B}$) are instead stored respectively on cell's edges and faces. \texttt{Spritz} evolves the magnetic field as the curl of a given vector potential $\bi{A}$, whose components are staggered just like the electric field $\bi{E}$ (see \Fref{Fig1}) and are usually evolved using the generalized Lorenz gauge. The electric and magnetic field components are not evolved variables. The electric field is computed at cell's edges using the flux-CT approach as described in the original WhiskyMHD paper~\cite{giacomazzo2007whiskymhd}. The magnetic field is instead computed at the cell faces using the vector potential component stored at cell's edges and then linearly interpolated at the center of the cell. The precise storage location on the grid--cells of various quantities is reported in \Tref{tabstagg}.

\begin{table}[t!]
\caption{\label{tabstagg}Location over the grid of various quantities. Symbols in the left column should be considered at the code's array position $\left( i,j,k \right)$ while the right column indicates the actual location over the grid that depends on whether the different quantities present a particular specification for the prolongation (for the components of the four--vector potential) or how they are computed within the code (for the components of the magnetic field).}
\footnotesize
\begin{center}
\begin{tabular}{@{}ccc}
\br
Symbol&Definition&Location\\
\mr
$\alpha$&lapse&$(i,j,k)$\\
$\beta^m$&$m$--component of the shift vector&$(i,j,k)$\\
$\gamma^{mn}$&$mn$--component of the spatial metric&$(i,j,k)$\\
$\gamma$&determinant of the spatial metric&$(i,j,k)$\\
$\rho$&rest-mass density&$(i,j,k)$\\
$p_{gas}$&pressure&$(i,j,k)$\\
$\varepsilon$&energy density&$(i,j,k)$\\
$v_m$&$m$--component of fluid velocity&$(i,j,k)$\\
$B^1$&$x$--component of magnetic field&$(i+\frac{1}{2},j,k)$\\
$B^2$&$y$--component of magnetic field&$(i,j+\frac{1}{2},k)$\\
$B^3$&$z$--component of magnetic field&$(i,j,k+\frac{1}{2})$\\
$A_1$&$x$--component of vector potential&$(i,j+\frac{1}{2},k+\frac{1}{2})$\\
$A_2$&$y$--component of vector potential&$(i+\frac{1}{2},j,k+\frac{1}{2})$\\
$A_3$&$z$--component of vector potential&$(i+\frac{1}{2},j+\frac{1}{2},k)$\\
$\Psi_\mathrm{mhd}$&scalar potential&$(i+\frac{1}{2},j+\frac{1}{2},k+\frac{1}{2})$\\
\br
\end{tabular}\\
\end{center}
\end{table}

Since $\bi{B}$ is computed from the curl of $\bi{A}$, the divergence--free character of the magnetic field is automatically satisfied.

The \texttt{Spritz} code evolves the vector potential $A$ and, when employing the generalized Lorenz gauge, the scalar potential $\Psi_\mathrm{mhd}$ is also computed.
Following \cite{etienne2015illinoisgrmhd}, we can write the update terms for the vector potential's components and for the scalar potential as follows:
\begin{equation}
\label{Axrhs}
\eqalign{
\partial_t A_m &= -E_m - \partial_m \left( G_{A} \right) = \cr
&= -E_m - \partial_m \left( \alpha \frac{\Psi_\mathrm{mhd}}{\sqrt{\gamma}} - \beta^j A_j \right),}
\end{equation}
for $m = 1$, $2$ and $3$, and
\begin{equation}
\label{Psirhs}
\eqalign{
\partial_t \Psi_\mathrm{mhd} &= -\partial_j \left( {F_{\Psi}}^j \right) - \xi \alpha \Psi_\mathrm{mhd} = \cr 
&= -\partial_j \left( \alpha \sqrt{\gamma} A^j - \beta^j \Psi_\mathrm{mhd} \right) - \xi \alpha \Psi_\mathrm{mhd},}
\end{equation}
being $\xi$ the so--called damping factor, used for the generalized Lorenz gauge. As the reader may deduce from \Eref{Axrhs}, \Eref{Psirhs}, and \Tref{tabstagg}, the terms on the right--hand sides in general have different storage locations and therefore we decided to follow this scheme:
\begin{enumerate}
\item At first we consider functions ${F_{\Psi}}^j$ and $G_{A}$, defined via \Eref{Axrhs} and \Eref{Psirhs} respectively, to be computed at cell centers, by interpolating $\Psi_\mathrm{mhd}$ and $A_j$ respectively from cell vertices and edges to the center.

\item Then we interpolate the values obtained at point (i) for ${F_{\Psi}}^j$ back to the cell edges and for $G_{A}$ back to cell vertices.

\item Finally we numerically differentiate the values at point (ii) via a centered difference scheme. For example, the derivative along $x$ ($m=1$) of $G_{A}$ in~\Eref{Axrhs} on the edge $(i,j+1/2,k+1/2)$ is computed as $[G_{A}(i+1/2,j+1/2,k+1/2)-G_{A}(i-1/2,j+1/2,k+1/2)]/(\Delta x)$. A similar expression is used for the derivatives computed at the cell vertex $(i+1/2,j+1/2,k+1/2)$ in ~\Eref{Psirhs} where the two nearby edges are used.
\end{enumerate}

In details, if a variable $f$ is given at cell vertices, then we interpolate it at the center of the cell using a simple linear interpolation:
\begin{equation}
\label{InterpVertices2Center}
\eqalign{
f(i,j,k) =& \frac{1}{8} \left[  f\left(i-\frac{1}{2},j-\frac{1}{2},k-\frac{1}{2}\right) + f\left(i-\frac{1}{2},j+\frac{1}{2},k-\frac{1}{2}\right) \right. \cr
&+ \left. f\left(i+\frac{1}{2},j+\frac{1}{2},k-\frac{1}{2}\right) + f\left(i+\frac{1}{2},j-\frac{1}{2},k-\frac{1}{2}\right) \right. \cr
&+ \left. f\left(i-\frac{1}{2},j-\frac{1}{2},k+\frac{1}{2}\right) + f\left(i-\frac{1}{2},j+\frac{1}{2},k+\frac{1}{2}\right) \right. \cr
&+ \left. f\left(i+\frac{1}{2},j+\frac{1}{2},k+\frac{1}{2}\right) + f\left(i+\frac{1}{2},j-\frac{1}{2},k+\frac{1}{2}\right) \right]}
\end{equation}
\Eref{InterpVertices2Center} is used to interpolate $\Psi_\mathrm{mhd}$ at step (i) of the aforementioned scheme.

For quantities defined instead on edges, for example along the $x$--direction, the following interpolation is instead used:
\begin{equation}
\label{InterpEdges2Center}
\eqalign{
f(i,j,k) =& \frac{1}{4} \left[  f\left(i,j-\frac{1}{2},k-\frac{1}{2}\right) + f\left(i,j+\frac{1}{2},k-\frac{1}{2}\right) \right. \cr
&+ \left. f\left(i,j-\frac{1}{2},k+\frac{1}{2}\right) + f\left(i,j+\frac{1}{2},k+\frac{1}{2}\right) \right]}
\end{equation}
\Eref{InterpEdges2Center} is used to interpolate $A_x$ at step (i) of the aforementioned scheme. Along other directions, the straightforward permutation of indices leads to the correct interpolating functions.

The following expression is instead used to interpolate from the cell center to a cell edge:
\begin{equation}
\label{InterpCenters2Edge}
\eqalign{
f\left(i,j+\frac{1}{2},k+\frac{1}{2} \right) =& \frac{1}{4} \left[ f(i,j,k) + f(i,j+1,k) \right. \cr
&+ \left. f(i,j,k+1) + f(i,j+1,k+1) \right]}  
\end{equation}
We use \Eref{InterpCenters2Edge} to obtain the values of ${F_{\Psi}}^j$ at point (ii). With the following interpolator we instead compute the values of $G_{A}$ at point (ii):
\begin{equation}
\label{InterpCenters2Vertex}
\eqalign{
f\left(i+\frac{1}{2},j+\frac{1}{2},k+\frac{1}{2}\right) =& \frac{1}{8} \left[  f(i,j,k) + f(i,j+1,k) \right. \cr
&+ \left. f(i,j,k+1) + f(i,j+1,k+1) \right. \cr
&+ \left. f(i+1,j,k) + f(i+1,j+1,k) \right. \cr
&+ \left. f(i+1,j,k+1) + f(i+1,j+1,k+1) \right]}
\end{equation}

In order to finally be able to compute the right-hand-side of \Eref{Axrhs}, one needs also to compute the electric field components $E_m$ that are stored at cell edges. Here we follow the same approach implemented in the \texttt{WhiskyMHD} code~\cite{giacomazzo2007whiskymhd} and use the flux-CT method~\cite{balsara1999staggered}, in which the electric field is computed from the magnetic field HLLE fluxes. Our staggered formulation therefore benefits of the same properties of the constrained transport scheme~\cite{evans88}, but without the need of implementing special prolongation and restriction operators to ensure the divergence-free character of the magnetic field~\cite{Balsara2001}.

\begin{figure}[hbt!]
\begin{center}
\includegraphics[width=0.5\linewidth]{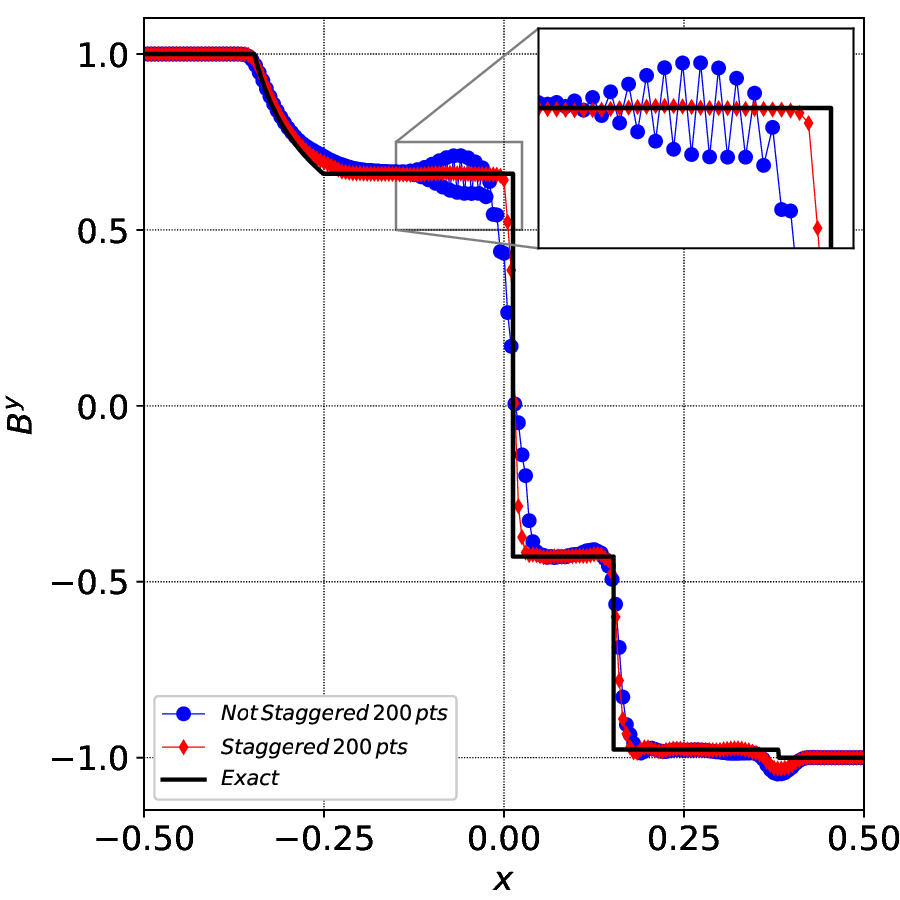}
\captionof{figure}{\label{Fig2}Comparison of results of \texttt{Balsara 1} test (from \cite{balsara2001total}), obtained via staggered (red diamonds) and non--staggered (blue dots) vector potential. Post--shock oscillations are clearly visible in the blue curve. It is worth noting that the non--staggered scheme is anyway still stable since the maximum amplitude of those oscillations does not grow indefinitely during the evolution. We remind that indeed \texttt{WhiskyMHD} applied a Kreiss-Oliger dissipation to the vector potential in order to remove such oscillations~\cite{giacomazzo2011accurate}.}
\end{center}
\end{figure}

An alternative scheme could use a non--staggered formulation where $\bi{A}$ and $\bi{B}$ are both stored at the cell centers (e.g., as done in the \texttt{WhiskyMHD} code~\cite{giacomazzo2011accurate}). An example of the different results for a shock--tube $1D$ test obtained via a staggered and a non--staggered scheme is shown in \Fref{Fig2}. 

\subsection{Boundary Conditions}
\label{BC}

When developing new codes to work within the \texttt{EinsteinToolkit} framework, the treatment of boundary conditions (BC) is usually left to the generic thorn \texttt{Boundary}~\cite{EinsteinToolkit:2019_10}. Through this approach, the \texttt{Spritz} code may consider ``flat'' or ``none'' BC, as already implemented in the \texttt{WhiskyMHD} \cite{giacomazzo2007whiskymhd} and \texttt{GRHydro} \cite{mosta2013grhydro} codes. The ``flat'' BC simply copies to the ghost zones the value that each variable has in the outermost grid point. The ``none'' BC instead does not update the ghost zones and keeps the value of the variables in the ghost zones equal to the ones set by the initial data routine.

Although the ``flat'' and ``none'' BC have been successfully used in simulations with the aforementioned codes, we decided to modify the BC at the external boundary of the computational domain for the vector and scalar potential in order to provide a more accurate calculation of B. We followed in particular the work presented in \cite{etienne2015illinoisgrmhd} and we implemented the numerical extrapolation of $\bi{A}$ and $\Psi_\mathrm{mhd}$ at the outer boundary as described in the \texttt{IllinoisGRMHD} code. Basically, for each grid--point in the outer boundary we apply the following linear extrapolation formula:
\begin{equation}
\label{numextrap}
F_{i}^{j} = \cases{2F_{i-1}^{j} - F_{i-2}^{j} \qquad \mbox{for } i = N^{j}-2,N^{j}-1,N^{j} \\
               2F_{i+1}^{j} - F_{i+2}^{j} \qquad \mbox{for } i = 3,2,1 }
\end{equation}
where $F \in \left\lbrace \bi{A}, \Psi_\mathrm{mhd} \right\rbrace$, $j \in \left\lbrace 1,2,3 \right\rbrace$, $N$ is the number of grid--points in the $j$--direction, and we use 3 points in the ghost zones for each direction.
In addition, the user may choose whether BC for $\bi{A}$ and $\Psi_\mathrm{mhd}$ should be given by \Eref{numextrap} or simply be obtained by the other two conditions provided by the \texttt{Boundary} thorn.

Finally, we also successfully tested the implementation of periodic BC provided by the thorn \texttt{Periodic}~\cite{EinsteinToolkit:2019_10} through the Loop Advection test (see \Sref{2D}), in both uniform and mesh--refined grids.

We note also that radiative BCs may be more suitable for GRMHD simulations, but these are not yet available in our code or in the Einstein Toolkit. Nevertheless, simulations of compact binary mergers already require large domains in order to compute GW signals. Therefore the effect of the BCs on the matter dynamics is negligible. Another approach, not yet implemented in the code, could be the use of multipatch methods such as those used in the \texttt{Llama} infrastructure~\cite{Pollney2011, Reisswig2013}.

\subsection{Primitive variables recovering}
\label{C2P}

As mentioned in \Sref{RieHRSC}, the computation of fluxes at each time during the evolution depends on values of the primitive variables $\bi{U}$, although we evolve the conserved ones $\bi{F}^0$.
As recurrent in many conservative approaches, one of the most delicate point is the inversion of \Eref{P2Csystem}, a problem that presents no analytic solution. Thus one has to apply a numerical method (usually a Newton-Raphson scheme).

In the literature many methods have been presented in order to perform this step~\cite{noble2006primitive, Siegel2018}. In the \texttt{Spritz} code we implemented both the 2D method used in \texttt{WhiskyMHD}~\cite{giacomazzo2007whiskymhd} and the one presented in \cite{noble2006primitive} and used in \texttt{GRHydro}.

\subsection{Atmosphere}
\label{Atmo}

As any GRMHD grid-based code, \texttt{Spritz} cannot handle zero values for the rest-mass density and a minimum value $\rho_\mathrm{atm}$ needs to be set. If at time $t$ the rest-mass density $\rho$ computed in our conservative-to-primitive routine is such that $\rho < \rho_\mathrm{atm}$, then its value is set to $\rho_\mathrm{atm}$, the pressure and specific internal energy are recomputed using a polytropic EOS, and the fluid's three--velocity is set to zero. In the tests presented here we typically set $\rho_\mathrm{atm} = 10^{-7} \rho_{0,\mathrm{max}}$, being $\rho_{0,\mathrm{max}}$ the initial maximum value of the rest-mass density.

\subsection{Equation of State}
\label{EOS}

To close the GRMHD system of equations, an equation of state that provides a relation between $\rho$, $\varepsilon$, and $p_\mathrm{gas}$ must be supplied. Many EOS exist, from analytical ones, such as that of an ``ideal fluid'' or of a ``polytropic'' gas~\cite{horedt2004polytropes}, to more complex ones that can only be expressed in a tabulated form~\cite{COMPOSE}. One of the most challenging research fields in astrophysics is focussed on trying to understand how matter behaves in the core of NSs, where the rest-mass density may reach values as high as $\sim\!10^{15}$ g cm$^{-3}$, not reproducible in Earth laboratories. Different EOS result in different bulk properties of the star, e.g., different maximum mass or equatorial radius for both spherical (i.e., non--rotating) and rapidly--rotating equilibrium configurations of NS models (see \cite{cipolletta2015fast} for examples taking into account EOS with various stiffness). It is therefore crucial for any GRMHD code to be able to handle different EOS with different composition as well as different treatments of nucleon interactions, in order to improve the capabilities of comparison between theoretical models and observations.

The \texttt{Spritz} code can implement both analytic and tabulated EOS. This is done via the \texttt{EOS\_Omni} thorn provided by the \texttt{EinsteinToolkit} which supports analytic EOS, such as ``ideal fluid'' and ``piecewise polytropic'' ones~\cite{Read2009}, and ``tabulated'' EOS.

For the sake of clarity, we report the explicit equations for the ``ideal fluid'' EOS, that can be written as
\begin{equation}
\label{IFEOS}
p_\mathrm{gas} = \left( \Gamma - 1 \right) \rho \varepsilon,
\end{equation}
where $\Gamma$ is the adiabatic index, and for the ``polytropic'' EOS, that reads
\begin{eqnarray}
\label{polyEOS}
p_\mathrm{gas} &=& K \rho^\Gamma\,,\\
\varepsilon &=& K \rho^{\Gamma-1}/(\Gamma - 1)\,,
\end{eqnarray}
being $K$ the polytropic constant. The tests presented in this paper will use only the ``ideal fluid'' EOS. A follow-up paper will present instead tests with cold and finite temperature equations of state, including also the evolution of the electron fraction and neutrino emission.

\subsection{Adaptive Mesh Refinement}
\label{AMR}

Adaptive Mesh Refinement (AMR) is very important in full 3D simulations of binary mergers because it allows for the optimization of the number of grid points by refining only interesting regions of the domain while maintaining a sufficiently large computational domain to reduce the effects of external boundaries and to allow for the extraction of gravitational wave signals far away from the source.

In the \texttt{EinsteinToolkit} framework~\cite{ETKpaper, EinsteinToolkit:2019_10}, this task is performed via the \texttt{Carpet} driver \cite{Carpet,schnetter2004evolutions}. Particular care must be taken in case of staggered variables, like $\bi{A}$ and $\Psi_\mathrm{mhd}$ in the \texttt{Spritz} code, as already mentioned in \Sref{Stagg}. In particular, one needs to use separate restriction and prolongation operators with respect to variables located at the cell centers. Such operators are already provided by the \texttt{Carpet} driver and they are the same used by the \texttt{IllinoisGRMHD} code. In \Sref{2D} we show also some tests of our AMR implementation.

\subsection{Spacetime Evolution}
\label{NumSTevo}

The spacetime evolution is performed using the \texttt{McLachlan} thorn~\cite{Brown:2008sb, Kranc:web, McLachlan:web}, which is part of the \texttt{EinsteinToolkit}. It adopts the BSSNOK formulation presented in \cite{baumgarte1998numerical,nakamura1987general,shibata1995evolution} and for which the numerical implementation has been presented in \cite{baiotti2005three,alcubierre2003gauge,alcubierre2000towards}. More details on the code can be found in~\cite{ETKpaper}.

\section{Results}
\label{sec4}

As already stressed in the Introduction, the central goal of the \texttt{Spritz} code is to perform simulations of BNS and NS-BH binary mergers. 
In order to address such a complex task with the necessary confidence, we need to assess the reliability of the code in a variety of physical conditions.
In this Section, we report on the results of our extensive testing, including a number of 1--, 2-- and 3--dimensional simulations. These simulations include critical tests that have been already considered in the literature in several previous papers (see, e.g., \cite{mosta2013grhydro,etienne2015illinoisgrmhd,balsara2001total,beckwith2011second} and references therein), allowing for a direct comparison with other codes.

\subsection{1D tests}
\label{1D}

\begin{figure}[hbt!]
\begin{center}
\includegraphics[width=\linewidth]{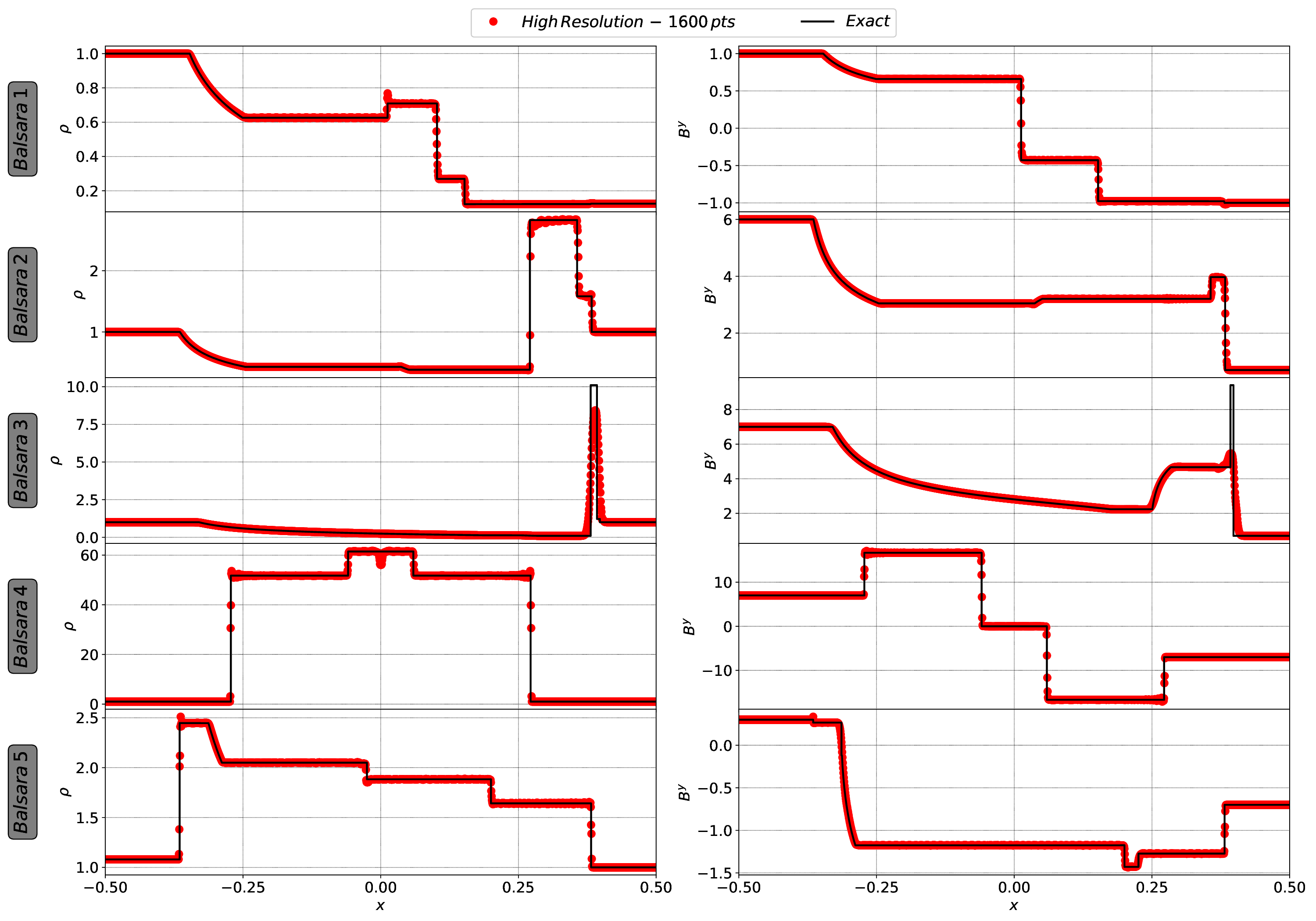}
\captionof{figure}{\label{Fig3}Comparison of numerical results (red dots) and exact solutions (continuous black lines) for the suite of tests of \cite{balsara2001total}. Left and right columns show respectively the spatial distributions of the rest-mass density and the magnetic field component $B^y$ at the final time of the evolution. 
Here the \texttt{Balsara 1}, \texttt{2}, \texttt{4} and \texttt{5} tests are performed with the third order \texttt{PPM} method. On the other hand, the \texttt{Balsara 3} is performed with the second order \texttt{MINMOD} method, because this test is the most demanding one due to the very high jump of four orders of magnitude in the initial pressure and this requires a slightly more dissipative method to succeed.}
\end{center}
\end{figure}
\begin{table}[t]
\footnotesize
\begin{center}
\begin{tabular}{@{}cD{.}{.}{4.1}D{.}{.}{4.1}D{.}{.}{4.1}D{.}{.}{4.1}D{.}{.}{4.1}D{.}{.}{4.1}D{.}{.}{4.1}D{.}{.}{4.1}D{.}{.}{4.1}D{.}{.}{4.1}D{.}{.}{4.1}D{.}{.}{4.1}D{.}{.}{4.1}D{.}{.}{4.1}D{.}{.}{4.1}D{.}{.}{4.1}}
\br
Test: & \multicolumn{2}{c}{ \texttt{1} } & \multicolumn{2}{c}{ \texttt{2} } & \multicolumn{2}{c}{ \texttt{3} } & \multicolumn{2}{c}{ \texttt{4} } & \multicolumn{2}{c}{ \texttt{5} } \\
\mr
& L & R & L & R & L & R & L & R & L & R \\
\mr
\lineup $\rho$ & 1.0 & 0.125 & 1.0   & 1.0 & 1.0       & 1.0  & 1.0     & 1.0      & 1.08 & 1.0   \\
\lineup $p_{gas}$     & 1.0 & 0.1     & 30.0 & 1.0 & 1000.0 & 0.1  & 0.1     & 0.1      & 0.95 & 1.0   \\
\lineup $v_x$  & 0.0 & 0.0     & 0.0   & 0.0 & 0.0       & 0.0  & 0.999 & -0.999 & 0.4  & -0.45 \\
\lineup $v_y$  & 0.0 & 0.0     & 0.0   & 0.0 & 0.0       & 0.0  & 0.0     & 0.0      & 0.3  & -0.2   \\
\lineup $v_z$  & 0.0 & 0.0     & 0.0   & 0.0 & 0.0       & 0.0  & 0.0     & 0.0      & 0.2  & 0.2    \\
\lineup $B^x$  & 0.5 & 0.5    & 5.0    & 5.0 & 10.0    & 10.0 & 10.0   & 10.0    & 2.0 & 2.0     \\
\lineup $B^y$  & 1.0 & -1.0   & 6.0    & 0.7 & 7.0      & 0.7   & 7.0     & -7.0    & 0.3  & -0.7   \\
\lineup $B^z$  & 0.0 & 0.0    & 6.0    & 0.7 & 7.0      & 0.7   & 7.0     & -7.0    & 0.3  & 0.5     \\
\br
\end{tabular}\\
\caption{\label{tab1D}Initial data for \texttt{Balsara} relativistic shock tube tests.}
\end{center}
\end{table}
\begin{figure}[hbt!]
\begin{center}
\includegraphics[width=\linewidth]{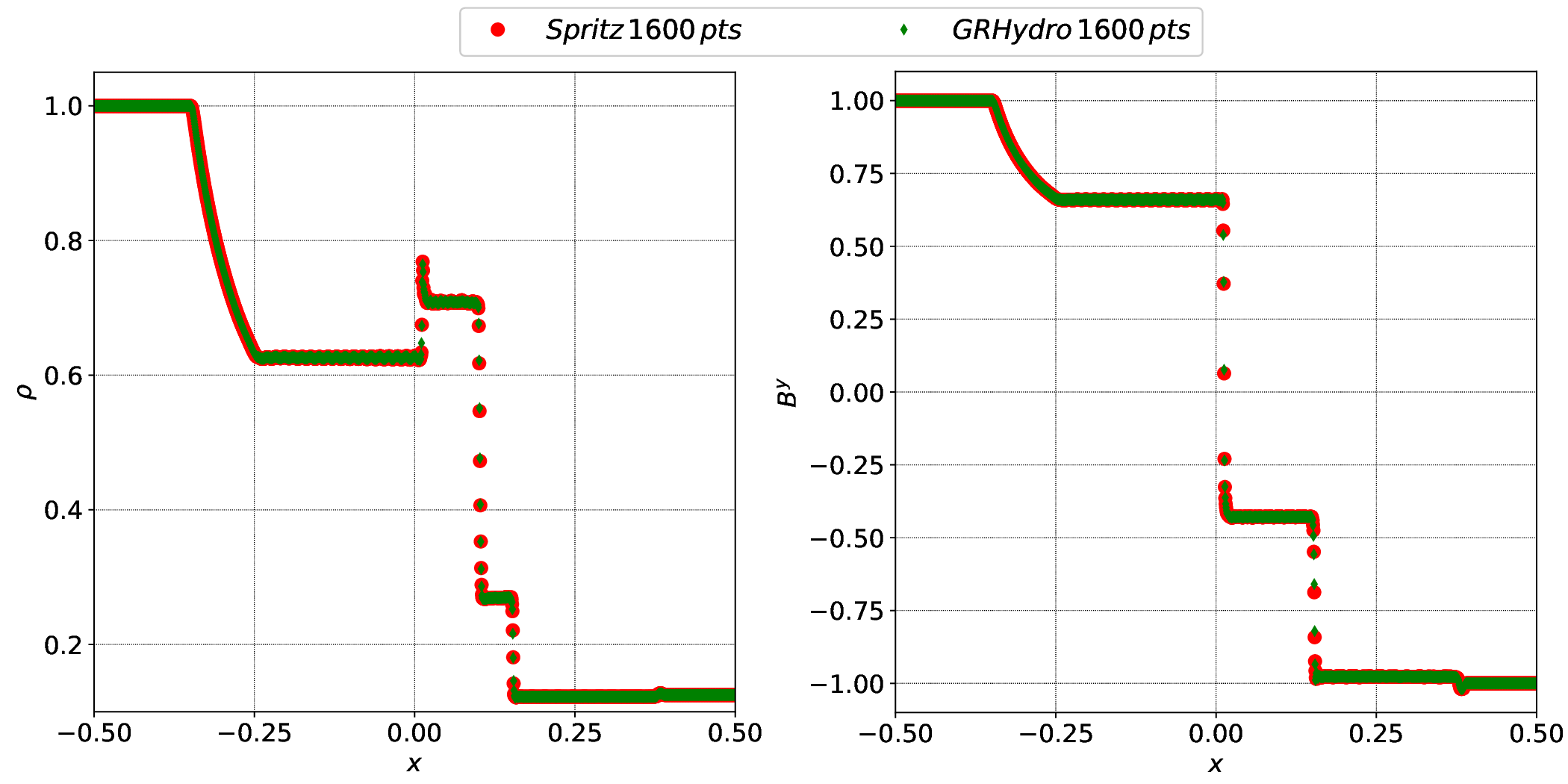}
\captionof{figure}{\label{Fig4}Comparison of results on the \texttt{Balsara 1} test (from \cite{balsara2001total}) obtained with the \texttt{Spritz} code (red dots) and the \texttt{GRHydro} code (green diamonds) \cite{mosta2013grhydro}.}
\end{center}
\end{figure}

The first tests that any GRMHD code should pass are those involving Riemann problems in order to check the correctness of the approximate Riemann solvers implemented in the code.
In \Fref{Fig3}, we present the results for 1--dimensional (1D) relativistic shock--tube problems corresponding to the suite of tests of \cite{balsara2001total}. Here, our numerical solution of such problems can be directly compared with the exact solutions computed via the code presented in \cite{giacomazzo2006exact}. Initial data for such tests are described in \Tref{tab1D}. In all tests we employ an ideal fluid EOS, with $\Gamma = 2.0$ for test \texttt{Balsara 1} and $\Gamma = 5/3$ for the others. The final evolution time is $t = 0.55$ for test \texttt{Balsara 5} and $t = 0.4$ for the others.
All tests show an excellent agreement between the numerical results and the exact solutions.

We also compared the results of these 1D tests obtained with the \texttt{Spritz} code with those already published for the numerical code \texttt{GRHydro} \cite{mosta2013grhydro}, finding a perfect match. 
In \Fref{Fig4}, we show an example of such comparison referring to the \texttt{Balsara 1} shock--tube test.

\begin{figure}[hbt!]
\begin{center}
\includegraphics[width=\linewidth]{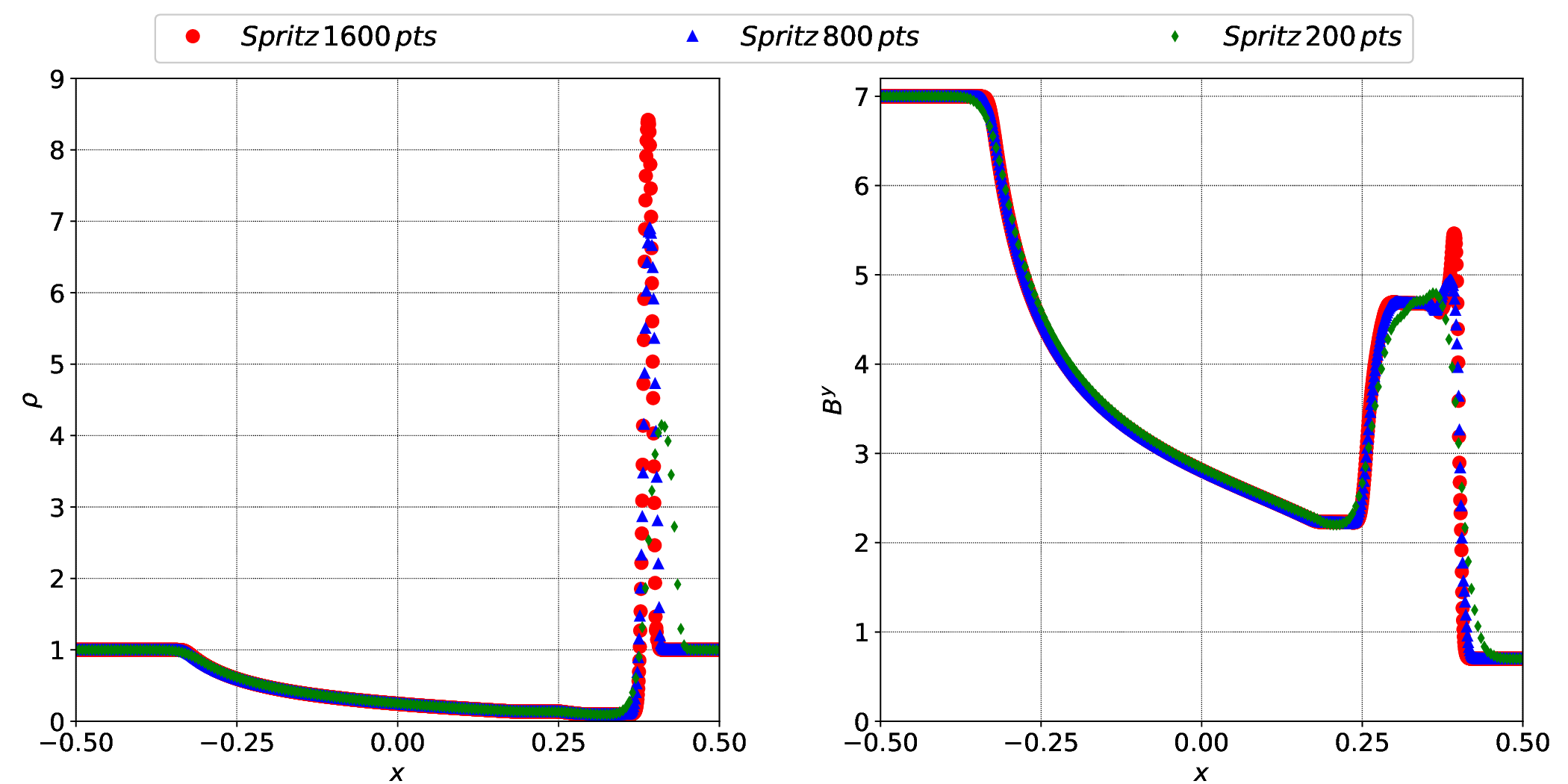}
\captionof{figure}{\label{Fig5}Comparison of results on the \texttt{Balsara 3} test~\cite{balsara2001total} obtained with the \texttt{Spritz} code at three different resolutions: low resolution (200 points -- green diamonds), medium resolution (800 points -- blue triangles) and high resolution (1600 points -- red dots).}
\end{center}
\end{figure}

Finally, \Fref{Fig5} shows our results on the most demanding \texttt{Balsara 3} test with different resolutions (200, 800, and 1600 grid points). Higher resolution leads to a significant increase in accuracy, which is particularly evident at the shock front (compare also with the exact solution in \Fref{Fig3}).

\subsection{2D tests}
\label{2D}
We now move on to discuss 2D tests performed with the Spritz code. 
In this work, we considered three types of 2D tests, namely the cylindrical explosion, the magnetic rotor and the magnetic loop advection, all performed in Cartesian coordinates. 
The cylindrical explosion test allows us to check the capability of the code to follow a shock front on the equatorial plane (such as for instance the ones that can be produced in a merging BNS system during the very last orbit prior to merger). The magnetic rotor test is a special relativistic test that, in a setup as simple as possible, allows to start testing the evolution of the magnetic field in the presence of rotation. Finally, the magnetic loop advection test is instead the only 2D test here with an exact solution to be compared with and as such it allows us also to better test the differences between our reconstruction schemes. We discuss all of them in some detail in the following subsections.

\subsubsection{Cylindrical Explosion}
\label{CylExp} 
\hfill\\

\noindent The cylindrical explosion (also known as the cylindrical blast wave) is a demanding multidimensional shock test, first introduced by \cite{komissarov1999godunov}, and later modified and implemented in \cite{mosta2013grhydro,etienne2010relativistic,beckwith2011second,del2007echo}. This test considers a uniformly magnetized domain consisting of a dense, over--pressured cylinder in the central region expanding in a surrounding ambient medium. Here, we adopt the parameters from the setup described in \cite{mosta2013grhydro}. For the cylinder, we set 
\begin{equation}
\label{CylBWintpar}
r_\mathrm{in}= 0.8, \; r_\mathrm{out}= 1.0, \; \rho_\mathrm{in} = 10^{-2}, \; p_\mathrm{gas,in} = 1.0 , 
\end{equation}
while for the surrounding ambient medium, we set 
\begin{equation}
\label{CylBWextpar}
\rho_\mathrm{out} = 10^{-4}, \; p_\mathrm{gas,out} = 3 \times 10^{-5}. 
\end{equation}

\noindent Here, $r_\mathrm{in}$ and $r_\mathrm{out}$ are the radial parameters used for the density profile smoothening prescription (and similarly for the pressure profile smoothening prescription) considered in \cite{mosta2013grhydro}, such that

\begin{equation}
\label{CylBWdens}
\rho(r) = \cases{ \rho_\mathrm{in} \qquad \qquad \qquad \qquad \qquad \qquad \qquad \quad \ ; \ r \leq  r_\mathrm{in}  \\
	\exp \Bigg[ \frac{(r_\mathrm{out} - r) \ln \rho_\mathrm{in} + (r-r_\mathrm{in}) \ln \rho_\mathrm{out}}{r_\mathrm{out} - r_\mathrm{in}}  \Bigg] \, ; \ r_\mathrm{in} < r <  r_\mathrm{out} \\ 
	\rho_\mathrm{out} \qquad \qquad \qquad \qquad \qquad \qquad \qquad \quad ; \ r \geq  r_\mathrm{out}
}
\end{equation}

\noindent The fluid velocity is initially set to zero and the magnetic field is initially uniform with $B^x = 0.1$ and $B^y = B^z = 0$. The test is performed on a $200 \times 200$ grid with x-~and y-coordinates spanning over the range $[-6,6]$. We adopt a Courant factor of 0.25 and an ideal fluid EOS with adiabatic index $\Gamma = 4/3$. We use the second order MINMOD reconstruction method along with the HLLE flux solver and the RK4 method for time-step evolution. 

\begin{figure}[hbt!]
\begin{center}
\includegraphics[width=0.9\linewidth]{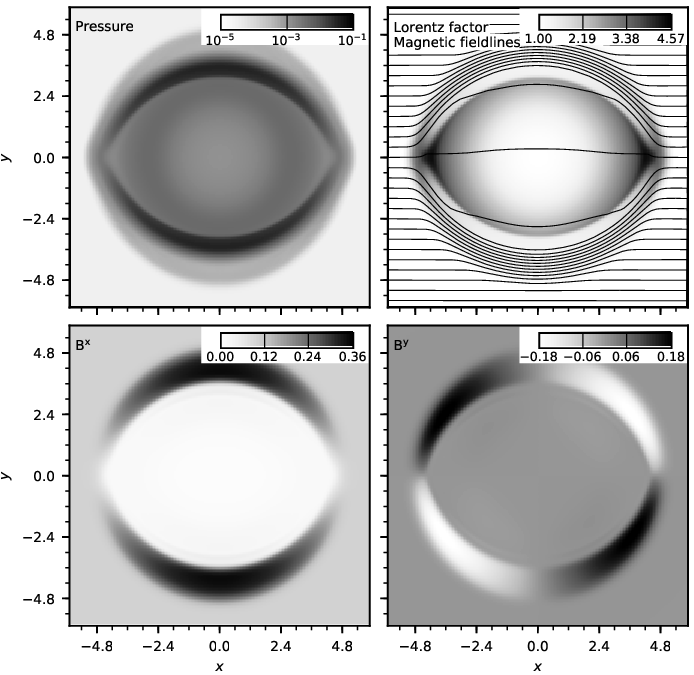}
\captionof{figure}{\label{Fig6} Cylindrical explosion test snapshots at the final evolution time $t=4$, showing the distribution of gas pressure $p_\mathrm{gas}$ (top--left), Lorentz factor $W$ together with magnetic field lines (top--right), and x-~and y-components of the magnetic field, $B^x$ (bottom--left) and $B^y$ (bottom--right). The resolution considered here is $\Delta x = \Delta y = 0.06$.}
\end{center}
\end{figure}

The resulting structure of the blast wave is shown in \Fref{Fig6} for the final time $t=4.0$. In particular, we show the two-dimensional distribution of gas pressure $p_\mathrm{gas}$, Lorentz factor $W$ (together with magnetic field lines), and the $x$-- and $y$--components of the magnetic field, $B^x$ and $B^y$. This Figure shows a very similar behavior as compared to the results already presented in the literature \cite{mosta2013grhydro,etienne2010relativistic,del2007echo}. 

\begin{figure}[hbt!]
\begin{center}
\includegraphics[width=0.95\linewidth]{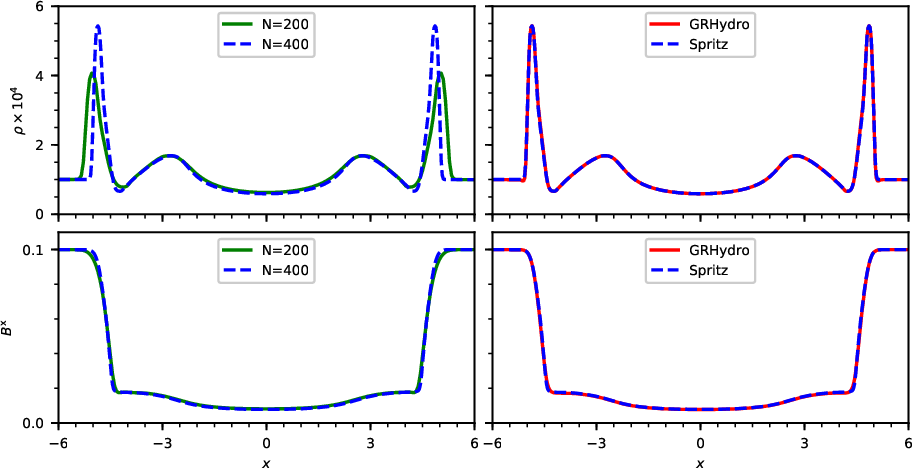} 
\captionof{figure}{\label{Fig7} One-dimensional cut along the x-axis of the cylindrical explosion test for the final evolution time $t=4$. A comparison between a low resolution test with $N=200$ grid--points (green solid line) and high resolution test with $N=400$ grid--points (blue dashed line) is shown in the left--side panels for the rest-mass density $\rho$ (top) and $x$--component of the magnetic field $B^x$ (bottom). The right--side panels show a comparison of the same quantities between the high resolution test performed with \texttt{Spritz} (blue dashed line) and the same test performed with \texttt{GRHydro} (red solid line).}
\end{center}
\end{figure}

\begin{figure}[hbt!]
\begin{center}
\includegraphics[width=0.9\linewidth]{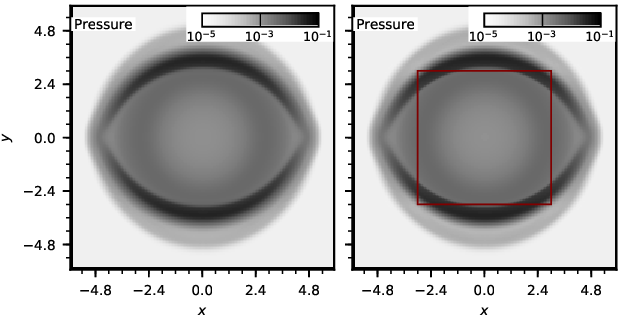}
\captionof{figure}{\label{Fig8} Cylindrical blast wave test with adaptive mesh refinement (AMR). A comparison is made for the final pressure configuration at $t=4$ between the low resolution test (left panel) performed with uniform grid and spacing $\Delta x = \Delta y = 0.06$ and the AMR test (right panel) including a refined inner grid (red box) with double resolution (i.e., grid-spacing $\Delta x = \Delta y = 0.03$). The two results are in agreement and no spurious effects are observed at the inner grid boundary.}
\end{center}
\end{figure}

\Fref{Fig7} provides instead a quantitative indication of the accuracy of our code. In this case, we show a one-dimensional slice along $y=0$ of the final blast wave configuration at time $t=4.0$, in terms of rest-mass density and $B^x$. 
Two cases are considered: on the left, we compare the results obtained with low ($200$ grid--points) and high resolution ($400$ grid--points); on the right, we compare the high resolution test results obtained with \texttt{Spritz} with those obtained with \texttt{GRHydro} \cite{mosta2013grhydro}. 
For the first comparison, we notice that the peaks differ slightly in $\rho$ due to the fact that the ability to capture the peak sharpness depends significantly on resolution. The values of $B^x$ show a much smaller deviation, due to a smoother variation of this quantity. 
For the second comparison, the agreement between \texttt{Spritz} and \texttt{GRHydro} appears excellent, further verifying the robustness of our code.

To validate the implementation of adaptive mesh refinement, we carried out another simulation including an inner refined grid covering the x-~and y-coordinates in the range [-3,3] with grid-spacing $\Delta x = \Delta y = 0.03$ (while the rest of the domain has double grid spacing). 
\Fref{Fig8} shows the comparison with the uniform grid test in terms of final pressure distribution. No significant differences are found, nor specific effects at the inner grid separation boundary, demonstrating a correct implementation of the AMR infrastructure.

\subsubsection{Magnetic Rotor} 
\label{MagRot}
\hfill\\

\noindent The second two-dimensional test we consider is the magnetic cylindrical rotor, originally introduced for classic MHD in \cite{balsara1999staggered,toth2000b} and later employed also for relativistic MHD in \cite{etienne2010relativistic,del2003efficient}. The initial setup of this test consists of a dense, rapidly spinning fluid at the center, surrounded by a static ambient medium, where the entire domain is set with a uniform magnetic field and pressure. For setting the initial parameters, we take the radius of the inner rotating fluid as $r=0.1$, with inner rest-mass density $\rho_\mathrm{in} = 10.0$, uniform angular velocity $\Omega=9.95$, and therefore the maximum value of the fluid three--velocity is $v_\mathrm{max} = 0.995$. For the outer static ambient medium, we set the rest-mass density as $\rho_\mathrm{out} = 1.0$. The initial magnitudes of the magnetic field and gas pressure are $B^i=(1.0, 0, 0)$ and $p_{gas,\mathrm{in}} = p_{gas,\mathrm{out}} = 1.0$. The problem is set up on a $400 \times 400$ grid with $x$-- and $y$--coordinates lying in range $[0,1]$. Here, we fix the Courant factor to 0.25 and consider an ideal fluid EOS with adiabatic index $\Gamma = 5/3$. For the system evolution, we use the second order MINMOD reconstruction method, the HLLE flux solver, and the RK4 method for time--stepping.

\begin{figure}[hbt!]
\begin{center}
\includegraphics[width=0.9\linewidth]{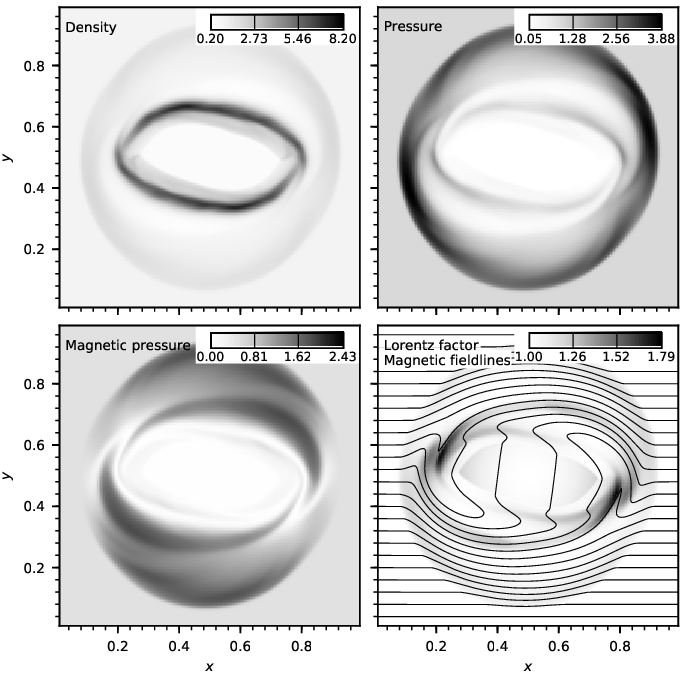}
\captionof{figure}{\label{Fig9} Magnetic rotor test with the following parameters shown for final evolved time $t=0.4$: Density $\rho$ (top--left), Gas pressure $p_\mathrm{gas}$ (top--right), Magnetic pressure $p_\mathrm{mag}$ (bottom--left),and Lorentz factor $W$ together with magnetic field lines (bottom--right). The resolution considered is $\Delta x = \Delta y = 0.0025$.}
\end{center}
\end{figure}

\Fref{Fig9} shows the two-dimensional profiles of density $\rho$, gas pressure $p_\mathrm{gas}$, magnetic pressure $p_\mathrm{mag}= b^2/2$, and Lorentz factor $W$ along with magnetic field lines, all at the final time $t=0.4$. The rotation of the cylinder causes magnetic winding. As one can see in the bottom--right panel of \Fref{Fig9}, the field lines are twisted roughly by $\sim 90^\circ$ in the central region. This twisting of field lines eventually slows down the rotation of the cylinder. There is also a decrease in $\rho$, $p_\mathrm{gas}$, and $p_\mathrm{mag}$ in the central region, observed along with the formation of an oblate shell of higher density. Also for this test, the results are in good agreement with the ones in the literature \cite{mosta2013grhydro,etienne2010relativistic,del2003efficient}.

\begin{figure}[hbt!]
\begin{center}
\includegraphics[width=0.95\linewidth]{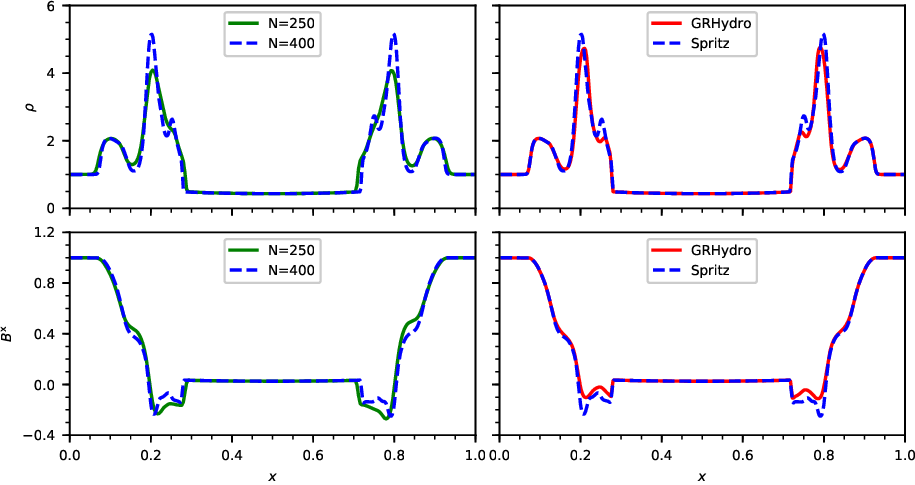}
\captionof{figure}{\label{Fig10} One--dimensional cut along the $x$--axis for the magnetic rotor test at the final evolution time $t=0.4$. A comparison between a low resolution test with $250$ grid--points (green solid line) and a high resolution test with $400$ grid--points (blue dashed line) is shown for the rest-mass density $\rho$ and the $x$--component of the magnetic field $B^x$ in the top and bottom left--side panels, respectively. In the top and bottom right--side panels, the same quantities are compared for the analogous high resolution test performed with \texttt{Spritz} (blue dashed line) and with \texttt{GRHydro} \cite{mosta2013grhydro} (red solid line).}
\end{center}
\end{figure}

Similarly to the test discussed in \sref{CylExp}, we perform a quantitative check by taking a one--dimensional slice along $y=0$ of the final rotor configuration at $t=0.4$. Again two cases are considered: (i) results comparison for the low and high resolution runs having $250$ and $400$ grid--points, respectively; (ii) results comparison for our high resolution test with the corresponding one preformed with \texttt{GRHydro} \cite{mosta2013grhydro}. 
\Fref{Fig10} shows this comparison made for the two quantities $\rho$ and $B^x$. For (i), as the resolution is increased, the peaks in $\rho$ as well as $B^x$ are better captured, showing signs of convergence towards the expected solution. For (ii), except for a minor difference in the peak values, the curves are comparable.  

\subsubsection{Loop Advection}
\label{LoopAdv}
\hfill\\

\noindent The third and last two--dimensional test we performed is the advection of a magnetic field loop, which was first described in \cite{devore1991flux} and appeared later in a slightly modified version (the one we consider) in \cite{mosta2013grhydro,beckwith2011second,gardiner2005unsplit,stone2008athena}. In this test, a magnetized circular field loop is propagated within the surrounding non--magnetized ambient medium with a constant velocity in a two--dimensional periodic grid. In particular, the analytical prescription for the initial imposed magnetic field (taken from \cite{mosta2013grhydro}) is given by
\begin{equation}
\label{LoopAdvMag}
B^x, \ B^y = \cases{ -A_\mathrm{loop}y/r, \ A_\mathrm{loop}x/r \ ; \quad r<R_\mathrm{loop} \\
	\qquad \qquad \quad \qquad \ \; \; 0 \ ; \quad r\geq R_\mathrm{loop}
}
\end{equation}
where $A_\mathrm{loop}$ is the amplitude of the magnetic field, $r = \sqrt{x^2 + y^2}$ is the radial coordinate, $R_\mathrm{loop}$ is the loop radius, and $B^z$ is set to zero. The corresponding vector potential prescription from which \Eref{LoopAdvMag} can be obtained is given by $\bi{A}(r) = (0,0,\mathrm{max}[0,A_\mathrm{loop}(R_\mathrm{loop}-r)])$~\cite{gardiner2005unsplit}.

For the initial parameters, we set the density as $\rho=1.0$ and pressure as $p_\mathrm{gas}=3.0$ throughout the computational domain. For the loop, we assume $A_\mathrm{loop}=0.001$ and $R_\mathrm{loop}=0.3$. The fluid 3-velocity is set to $v^i=(1/12, 1/24, 0)$ for a case where $v^z=0$ and $v^i=(1/12, 1/24, 1/24)$ for a more generic case in which the vertical component of the velocity is non-zero, i.e., $v^z\neq 0$. We run the test in both low resolution with a $128\times 128$ grid and high resolution with a $256\times 256$ grid, where the $x$-- and $y$--components span the range [-0.5,0.5]. The Courant factor is 0.4 and the adiabatic index for the ideal EOS is $\Gamma=5/3$.  Like the previous 2D tests, we utilize the MINMOD reconstruction method along with the HLLE flux solver and the RK4 method for time-step evolution. 

\begin{figure}[hbt!]
\begin{center}
\includegraphics[width=0.85\linewidth]{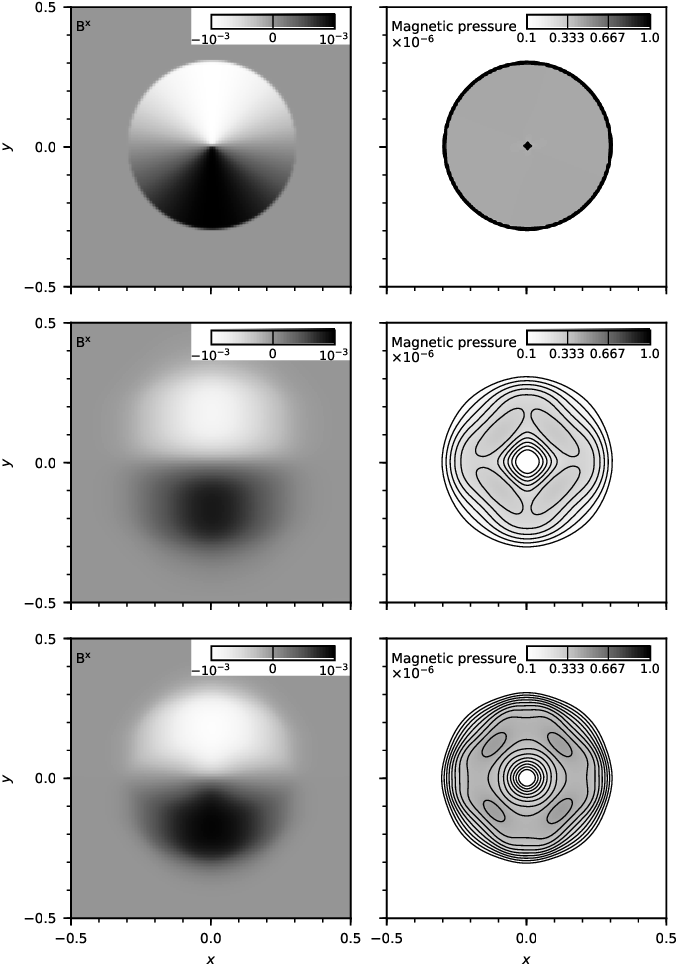}
\captionof{figure}{\label{Fig11} Loop advection test with $v^z=1/24$. Left and right columns represent the x-component of the magnetic field $B^x$ and the magnetic pressure $p_\mathrm{mag} = b^2/2$, respectively. The initial data for $B^x$ and its corresponding $p_\mathrm{mag}$ at $t=0$ is depicted in the top row, while middle and bottom rows represent these quantities after one periodic cycle of evolution, i.e., at $t=24$, in low resolution ($\Delta x = 1/128$) and high resolution ($\Delta x = 1/256$), respectively. Our results are in very good agreement with those reported in \cite{mosta2013grhydro}. }
\end{center}
\end{figure}

The outcome of the $v^z\neq 0$ test case is shown in \Fref{Fig11}. Here, the top row illustrates the initial configuration of the magnetic loop for the quantities $B^x$ and $p_\mathrm{mag}=b^2/2$ at $t=0$. After one entire cycle of the loop across the domain at $t=24$, the same quantities are depicted in the middle row for low resolution run and the bottom row for high resolution run. We notice a significant loss of magnetic pressure due to numerical dissipation for the low resolution test after one evolution cycle as also reported in \cite{mosta2013grhydro}, which is however smaller for higher resolution. Our results are comparable with the ones presented in \cite{mosta2013grhydro}. It is worth noting that the expression for magnetic pressure used for \Fref{Fig11} is $p_\mathrm{mag}=b^2/2$ and differs from the expression used for figure 10 of \cite{mosta2013grhydro} by a factor of $1/2$ (in~\cite{mosta2013grhydro} the authors actually plotted $b^2$).

\begin{figure}[hbt!]
\begin{center}
\includegraphics[width=\linewidth]{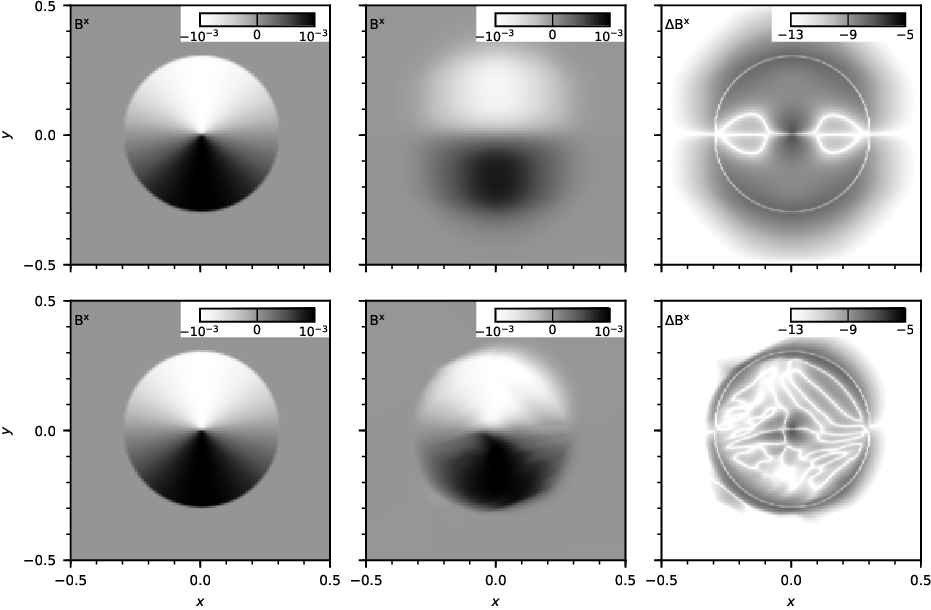}
\captionof{figure}{\label{Fig12} Comparison between the MINMOD and PPM reconstruction methods for the loop advection test with $v^z=1/24$. Top and bottom rows correspond to results obtained with MINMOD and PPM respectively. First column depicts the initial configuration of the magnetic field $B^x$ at $t=0$, second column shows the final configuration of $B^x$ after one periodic cycle at $t=24$, and the third column shows the logarithmic absolute differences in $B^x$ between the initial and final times.}
\end{center}
\end{figure}

To consider a less dissipative numerical scheme, we also perform another run in low resolution employing the PPM reconstruction and compare the results with those obtained with MINMOD reconstruction. This is shown in \Fref{Fig12}, where the top and bottom panels represent the outcome of the runs with MINMOD reconstruction and PPM reconstruction, respectively. 
The first column depicts the initial data at $t=0$, the second column shows the loop at final time $t=24$, while the third column shows the logarithmic values of the absolute differences between the initial and final times.
As expected, we find significantly lower dissipation in the PPM case. 

\subsection{3D tests}
\label{3D}

We now present the results of our 3D tests, mostly including a fully dynamical spacetime. The first one is the generalization in 3D of the cylindrical explosion test and it allows us to check the robustness of our code to handle 3D spherical shock waves (similar to those that can be formed for example during the merger of two NSs). The other tests instead follow the evolution of non-rotating NSs and these are the first tests in full general relativity, hence assessing the correctness of our code also in a curved and dynamical background.

\subsubsection{Spherical Explosion}
\label{SE}
\hfill\\

\noindent We present here the results of a very demanding GRMHD test which is not usually performed by other GRMHD codes and that is successfully passed by the \texttt{Spritz} code: the so--called Spherical Explosion.

Usually, GRMHD codes based on Cartesian coordinates are tested with the Cylindrical Explosion test (refer to \sref{CylExp}), because the cylindrical symmetry can be well exploited in such a geometrical setting. Spherical Explosion tests, instead, have commonly been performed with GRMHD codes working in spherical coordinates \cite{cerda2008new,cerda2007general,Mewes2020}, which are not well-suited for dealing with cylindrical symmetry. 
What make the Spherical Explosion test challenging in Cartesian coordinates are indeed the potential limitations in regions where the shock front is not parallel to the orientation of grid--cells' faces.

The test settings are an extension in spherical symmetry of the Cylindrical Explosion test of \sref{CylExp}. We consider an inner dense sphere of radius $R_\mathrm{in} = 0.8$ centered in the domain's origin with $\rho_\mathrm{in} = 10^{-2}$ and $p_{gas,\mathrm{in}} = 1.0$, surrounded by a spherical shell covering the radial range $R_\mathrm{in} < r < R_\mathrm{out} = 1.0$ where pressure and density are characterized by an exponential decay analogous to the prescription given in \Eref{CylBWdens}, except that here a spherical radius is considered instead of a cylindrical one. 
At $r > R_\mathrm{out}$, we have then a low-pressure uniform fluid with $\rho_\mathrm{out} = 10^{-4}$ and $p_{gas,\mathrm{out}} = 3.0 \times 10^{-5}$.  
In addition, following \cite{cerda2008new,cerda2007general}, a uniform magnetic field parallel to the z axis is added all over the domain. The domain extension is $ \left[ -6.0, 6.0 \right] $ and is covered by $160$ grid--cells, in all directions. Although a direct comparison with spherical coordinates settings of \cite{cerda2008new,cerda2007general} can not be done in a straightforward way, it is worth noting that this choice for the resolution corresponds to considering $80$ cells in the radial direction along the polar axis, i.e., the low--resolution version of the results presented in the aforementioned papers. We decided to perform the evolutions for a total time of $t_\mathrm{final} = 6.0$, with a CFL factor of $0.25$. 
Our runs did not crash even at this late time, although the shock--front always reaches the boundary of the domain (that is treated with ``none'' BCs). 
We also note that in this case the imposed lower limit for the rest-mass density (defining the atmospheric floor, see \Sref{Atmo}) is $\rho_\mathrm{atm} = 10^{-12}$.

\begin{figure}[t!]
\begin{center}
\includegraphics[width=0.8\linewidth]{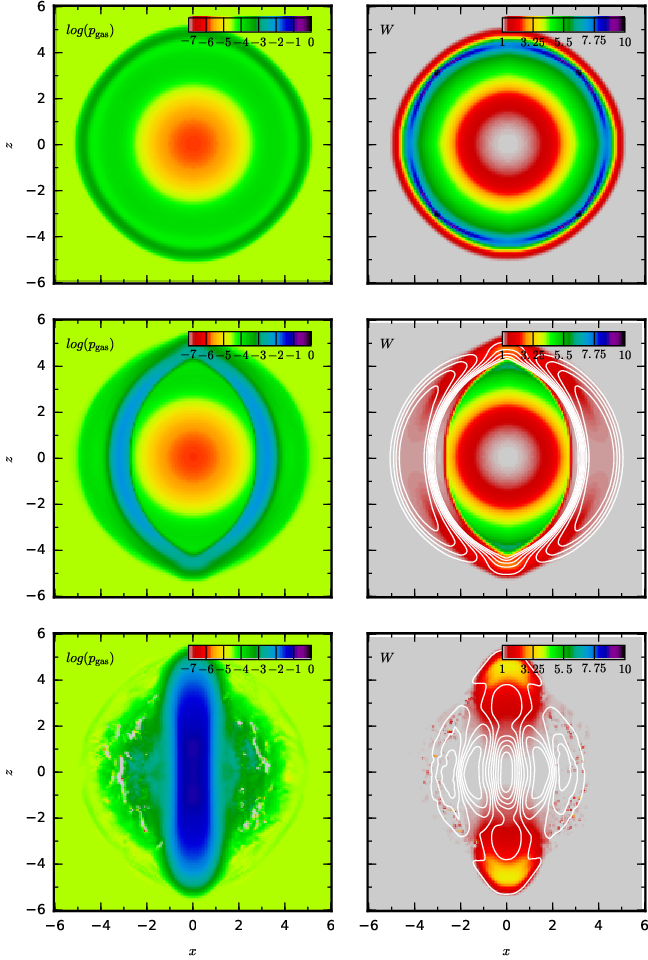}
\captionof{figure}{\label{Fig16} Spherical Explosion test results at time $t = 4.0$. Top row is the result for the non--magnetised case, middle row is the intermediate magnetization case with $B^z = 0.1$ and bottom row is the strongly magnetized case with $B^z = 1.0$. The left and right columns show respectively the logarithm of the gas pressure and the Lorentz factor along with isodensity contour lines of $ || B || = \sqrt{B^i B_i}$.}
\end{center}
\end{figure}

In \Fref{Fig16} we report on separated rows the results on the $y = 0$ plane of the tests performed respectively with magnetic field strength $B^z = 0.0, 0.1$ and $1.0$. In particular, we show the gas pressure and Lorentz factor $W$ (respectively on the left and right columns) at time $t=4$.
Looking a the top--right panel (Lorentz factor in the non--magnetized case), we can observe small deviations from spherical symmetry exactly aligned with the Cartesian axes, giving a hint of the geometrical issues brought by such a demanding test. 
In fact, as already noted by \cite{del2003efficient} for the Cylindrical Explosion, the biggest problems are due to the fluid velocity components along the diagonals. However, despite the accumulation of errors along the diagonals due to the non--perpendicularity of the fluxes, the spherical shape of the shock front seems to be very well preserved in this case, even at the relatively low resolution considered here. 

In presence of a dynamically important magnetic field oriented along the $z$ axis, the shock front deviates naturally from spherical symmetry (see middle row of \Fref{Fig16}). Finally, when the magnetic field strength is very high (see bottom row), the central region gets completely evacuated. Even in such an extreme case, the evolution is still performed without any problem.

A final important note is that all the tests for the Spherical Explosion here presented where performed with the minmod reconstruction and the LxF flux method, but without adopting any additional dissipation or ad-hoc fixes.

\subsubsection{TOV star}
\label{TOV}
\hfill\\

\noindent Static, spherically symmetric stars in general relativity are best described by the Tolman--Oppenheimer--Volkoff (TOV) equations \cite{oppenheimer1939massive,tolman1939static}. To further assess the stability and accuracy of our code, the next test we considered is the evolution of a non--rotating stable TOV configuration for both non--magnetised and magnetised cases. For the test setup, we adopt the model described in \cite{baiotti2005three} that we build using the TOVSolver thorn \cite{EinsteinToolkit:2019_10}. 
In particular, the initial TOV star configuration is generated using a polytropic EOS with adiabatic index $\Gamma=2.0$, polytropic constant $K=100$, and initial central rest-mass density $\rho=1.28\times10^{-3}$. 
We perform the evolution of this initial configuration adopting an ideal fluid EOS with the same value for $\Gamma$. For the magnetised version, we add the magnetic field to the computed TOV configuration using the analytical prescription of the vector potential $A_\phi$ given by
\begin{equation}
	\label{VecPot}
	A_\phi \equiv A_\mathrm{b} \varpi^2 {\rm max} \left( p - p_\mathrm{cut}, 0 \right)^{n_s} \ ,
\end{equation}
where $\varpi$ is the cylindrical radius, $A_\mathrm{b}$ is a constant, $p_\mathrm{cut}=0.04p_\mathrm{max}$ determines the cutoff when the magnetic field goes to zero inside the NS, with $p_\mathrm{max}$ corresponding to the initial maximum gas pressure, and $n_s=2$ sets the degree of differentiability of the magnetic field strength \cite{giacomazzo2011accurate}. The value of $A_b$ is chosen such that the maximum value of the initial magnetic field strength is set to $\approx1\times 10^{16} \ \mathrm{G}$. This generates a dipole-like magnetic field confined inside the NS and zero magnetic field outside. 

The non-magnetised tests are run on a uniform grid with $x$--, $y$-- and $z$--coordinates spanning over the range [0, 20] with low, medium and high resolution having $(32)^3$, $(64)^3$ and $(128)^3$ grid--cells respectively, and considering reflection symmetry with respect to every direction, i.e., the so--called octant symmetry. Furthermore, we perform two more tests for non-magnetised TOV NS in high resolution (i) employing the Cowling approximation (i. e. considering a fixed space--time) \cite{cowling1941non,lindblom1990accuracy,1969ApJ...158..997T} to check the accuracy of our code by evolving just the hydrodynamical equations on a static spacetime background, and (ii) implementing a mesh refinement composed by two nested boxes centered at the origin and extending up to $x,y,z=$20 and 40, respectively, both having $(128)^3$ grid--cells in each direction (therefore the inner box corresponds to the domain evolved in the unigrid run at high resolution while the outer box allows for a further out external boundary). As the \texttt{EinsteinToolkit} does not provide a way to handle reflection symmetry for staggered variables, we perform the magnetised TOV tests in low, medium and high resolution covering the entire domain with $x$--, $y$-- and $z$--coordinates lying in the range [-20, 20] (considering no reflection symmetries) having the same respective grid-spacing as that of the non-magnetised simulations. 
All the test cases are simulated for $10$~ms using the PPM reconstruction method, the HLLE flux solver, and the RK4 method for time stepping with a CFL factor of $0.25$.

\begin{figure}[t!]   
	\centering
	\begin{minipage}{0.8\textwidth}
		\centering
	    \includegraphics[width=1\linewidth]{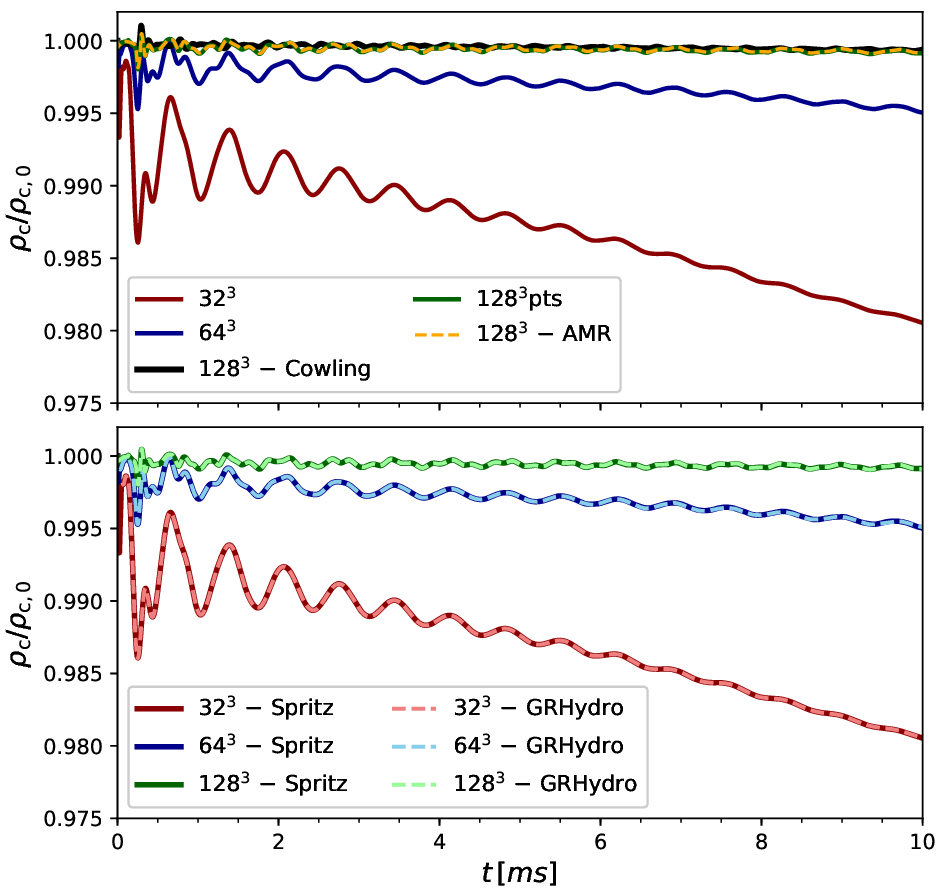} 
	\end{minipage}

\caption{Results of the non--magnetised TOV simulations. Top: Time evolution of the normalised central rest--mass density $\rho_\mathrm{c}/\rho_\mathrm{c,0}$ for the different resolution simulations inclusive of cases with Cowling approximation and AMR. Bottom: Comparison of results on the $\rho_\mathrm{c}/\rho_\mathrm{c,0}$ evolution with those obtained with \texttt{GRHydro} for low, medium, and high resolution, showing an exact match.}
\label{Fig13}
\end{figure}

The top panel of \Fref{Fig13} shows the central rest--mass density $\rho_\mathrm{c}$ evolution for all three resolutions, the high--resolution in Cowling approximation and the high--resolution with AMR, all for the non--magnetised TOV case. It is worth noting that the AMR case (orange curve) can reproduce perfectly the result in high--resolution (green curve), this proving once again the correctness of AMR implementation within the \texttt{Spritz} code. Periodic oscillations are initiated as a result of the truncation errors generated in the initial data, while the cause of dissipation is primarily due to the numerical viscosity of the finite differencing (FD) scheme \cite{baiotti2005three,font2000non}. The results converge well after increasing the resolution, and the additional tests for the cases with Cowling approximation and AMR are also fully satisfactory. In order to further investigate the accuracy of our code, we compare the low, medium, and high resolution tests' results on the $\rho_\mathrm{c}$ evolution with those obtained with \texttt{GRHydro}. As shown in the bottom panel of \Fref{Fig13}, we observe an exact match.

\begin{figure}[t!]
	\begin{center}
		\includegraphics[width=0.5\linewidth]{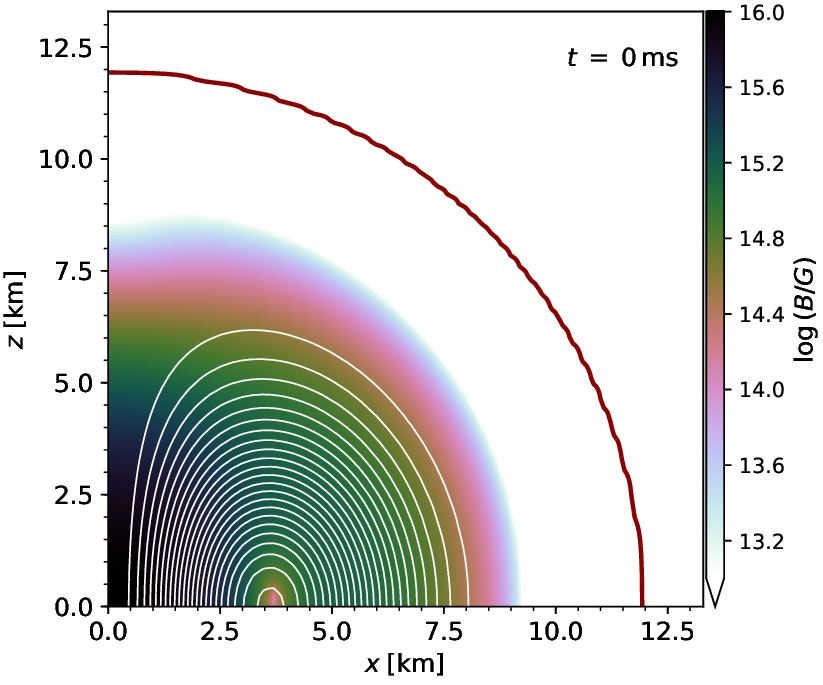}
        \caption{Initial internal magnetic field configuration of the magnetised TOV. The colormap indicates the strength of the magnetic field, while the contours (in white) trace a number of representative isosurfaces of the $\phi$--component of the vector potential, $A_\phi$. The latter contours also correspond to poloidal magnetic field lines. The red line is an approximate representation of the TOV surface, showing the iso-density contour of $5\times10^5$ times the assumed atmospheric floor density.} 
        {\label{Fig14} }
	\end{center}
\end{figure}

The initial magnetic field configuration for the magnetised TOV test is illustrated in \Fref{Fig14}. Here, the magnetic field strength is shown along with representative magnetic field lines.

\begin{figure}[t!]
	\centering
	\begin{minipage}{0.7\textwidth}
		\centering
		\includegraphics[width=1\linewidth]{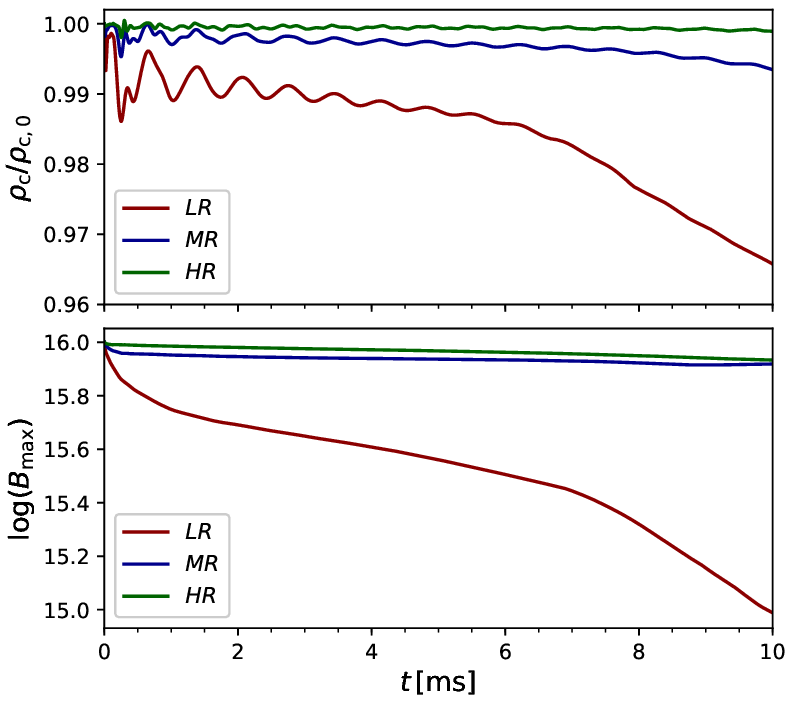} 
	\end{minipage}%

	\caption{Results of the magnetised TOV simulation. Top: Time evolution of the normalised central rest--mass density $\rho_\mathrm{c}/\rho_\mathrm{c,0}$ for the different resolution simulations; this gives a nearly exact match with that of the non-magnetised TOV case results (c.f., \Fref{Fig13}). Bottom: Time evolution of the maximum value of the magnetic field strength for all three resolutions.}
	\label{Fig15}
\end{figure}

The top panel of \Fref{Fig15} shows the evolution of the maximum of the rest-mass density $\rho_\mathrm{c}$ for the magnetised TOV case, which matches almost exactly the one for the non--magnetised case (see the top panel of \Fref{Fig13}). 
This should be expected, since the imposed magnetic field represents only a small perturbation compared to the gravitational binding energy of the system. 
In addition, the time evolution of the maximum value of the magnetic field strength $B_\mathrm{max}$ is depicted in the bottom panel of \Fref{Fig15}. While $B_\mathrm{max}$ is highly damped for the lowest resolution test with a decrease by a factor of roughly $14.75$ in $10$~ms, its value stabilizes with increasing resolution, as observed for $\rho_\mathrm{c}$. We note again that here the damping is a numerical viscosity effect of the FD scheme.

\begin{figure}[t!]
   	\centering
   \begin{minipage}{0.49\textwidth}
   	\centering
   	\includegraphics[width=1\linewidth]{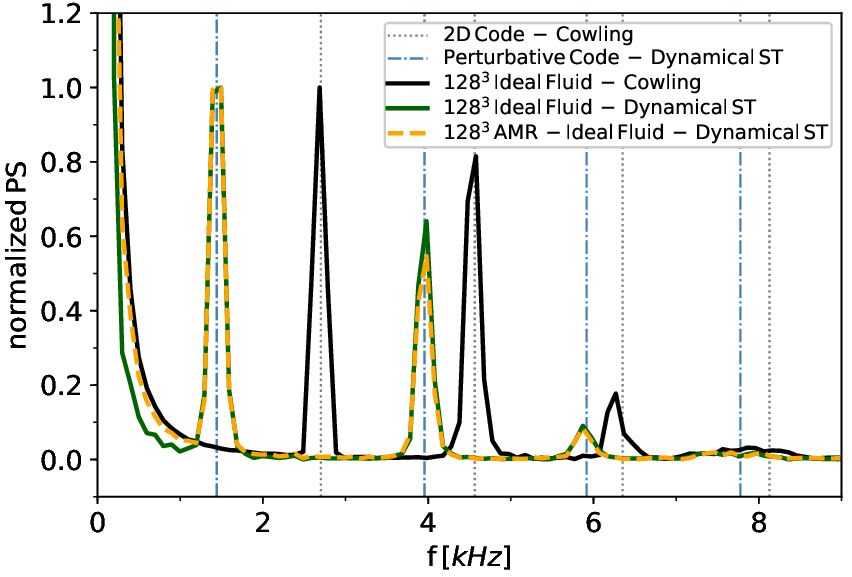} 
   \end{minipage}
   \begin{minipage}{0.49\textwidth}
     \centering
     \includegraphics[width=1\linewidth]{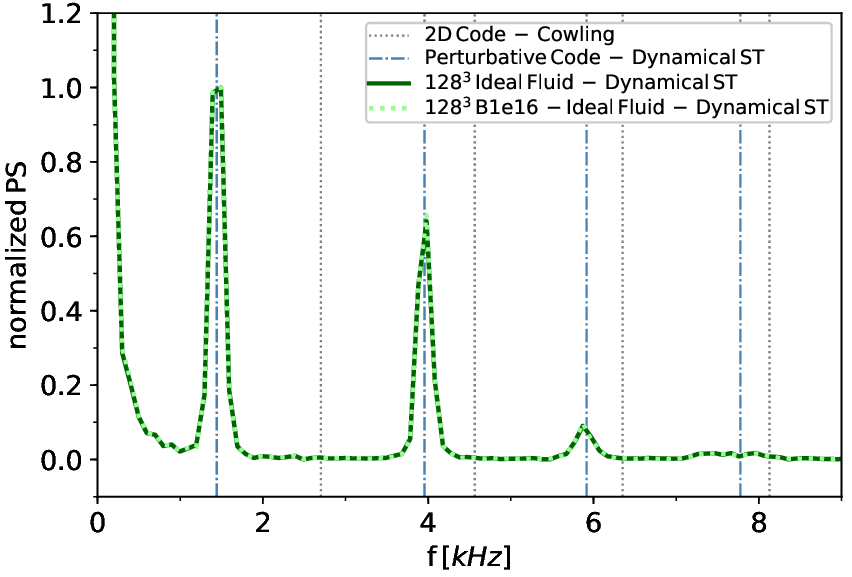} 
    \end{minipage}%
	\caption{Power spectrum of the central rest--mass density evolution, normalized to maximum amplitude of the peaks of oscillations' frequencies. Left--panel shows the results from the runs without magnetic field, while right--panel shows the results where also magnetic field is considered.}
	\label{Fig17}
\end{figure}

To conclude this section, we report in \Fref{Fig17} the oscillations' peak frequencies for the evolution of the TOV star models that were simulated with our code, in order to validate our models with the literature results. In particular, we show the results of the high--resolution simulations in pure hydrodynamics with dynamical space--time both adopting uniform grid and AMR, with the Cowling approximation (see \Fref{Fig17} left-panel), as well as of the high--resolution run with magnetic field (see \Fref{Fig17} right-panel). The power spectrum of each simulation is computed via fast Fourier transform (FFT) in order to extract the amplitudes and frequencies of the oscillations of the central rest mass density, and then the amplitudes are normalized to the maximum one relatively to each simulation. \Fref{Fig17} also shows the peak frequencies of the oscillations from the literature taken from \cite{font2002three}, that were obtained with an independent 2D code for fixed space--time and with a perturbative code in the case of hydrodynamics coupled to space--time evolution. An interesting point to note is that although the results of \cite{font2002three} were obtained with a polytropic EOS, our Ideal Fluid simulations seem to match perfectly the peak frequencies. The ideal fluid EOS produces indeed different results from a polytropic one only in presence of shocks, which in this case appears only on the low-density surface and therefore do not affect the oscillations of the core. Finally, it is worth noting that the peak frequencies of our non--magnetised and magnetised models are in perfect agreement as shown by the left panel of \Fref{Fig17}, proving the correctness of the magnetic field implementation.

\section{Conclusion and future developments}
\label{sec5}

We have presented a new fully general relativistic code, named  \texttt{Spritz}, able to evolve the GRMHD equations in three spatial dimensions, on cartesian coordinates, and on dynamical backgrounds. The code is based
and considerably improves over our previous \texttt{WhiskyMHD}
code~\cite{
  giacomazzo2007whiskymhd, giacomazzo2011accurate,GiacomazzoPerna2013}. The \texttt{Spritz} code benefits also from
the publicly available \texttt{GRHydro}~\cite{mosta2013grhydro} and
\texttt{IllinoisGRMHD}~\cite{etienne2015illinoisgrmhd} codes, in
particular in the handling of different EOSs and in the use of a
staggered formulation of the vector potential equations.

In this paper, we presented in detail the equations and the numerical
methods implemented in the code. We have adopted a conservative formulation of GRMHD equations, high-resolution shock-capturing schemes, and we guarantee the divergence-less character of the magnetic field by evolving the vector potential. We also presented a series of tests
in special and general relativity. We started by showing the code capability of accurately solving 1D Riemann problems by comparing the numerical results with exact solutions~\cite{giacomazzo2006exact}. We also showed, for the first time, a comparison between a
non-staggered and a staggered formulation of the vector potential,
demonstrating that the latter prevents the formation of spurious
post-shock oscillations (see~\Fref{Fig1}) and therefore does not
require to apply dissipation to the vector
potential~\cite{giacomazzo2011accurate}. We then performed a series of special relativistic MHD tests in 2D, including the cylindrical explosion, the magnetic rotor, and the loop advection tests. All tests showed very good agreement with the exact solution (loop advection) or with other GRMHD codes (cylindrical explosion and magnetic rotor). In the cylindrical explosion case we also tested the code capability of dealing with mesh refinement boundary and demonstrated that they have no effect in the correct evolution of MHD quantities. We also performed, for the first time for a fully GRMHD code, a demanding 3D spherical explosion test with different levels of magnetization. The code produced results in very good agreement with those produced by other codes. We concluded our series of tests with a standard 3D evolution of a stable TOV configuration (both with and without magnetic field) in order to show the code
ability to handle fully general relativistic regimes. In particular we checked the frequency of TOV oscillations and compared them with results available in the literature.

While the \texttt{Spritz} code can handle any equation of state, in this paper we focused on tests using simple gamma-law EOSs in order to check the robustness of our basic GRMHD routines. In a second paper we will present also tests involving the evolution
of isolated NSs with finite temperature EOSs and neutrino emission with and without magnetic fields (Sala et al., in preparation). 

Once this second family of tests will be performed successfully, the  \texttt{Spritz} code will be one of the very few codes worldwide able to evolve magnetised neutron stars with finite temperature EOSs and neutrino emission~\cite{most2019beyond,palenzuela2015effects}. In the
multimessenger era it is indeed crucial to take into account
different aspects of the microphysics in order to be able not only to
compute a more accurate merger and post-merge GW signal, but also
to provide reliable estimates of the EM emission, including both kilonova
and short GRBs. The former requires indeed an accurate estimate of electron fraction and temperature in the post-merger remnant as well as in the ejected material, while the latter needs a precise description of the magnetic field evolution.

The version of the code used for this paper is available for download from Zenodo~\cite{zenodo-spritz}. Once the \texttt{Spritz} code will have been used for a first set of binary NS merger simulations, we will also ask for its inclusion in future releases of the Einstein
Toolkit~\cite{ETKpaper, EinsteinToolkit:2019_10}.

\hfill\\
\section*{Acknowledgments}
Numerical calculations have been made possible through a CINECA-INFN agreement, providing access to resources on MARCONI at CINECA. F. C. acknowledges financial support from the INFN HPC\_HTC project. F. C. acknowledges the CCRG at the RIT for the computational resources provided there on the \textit{Green Prairies} local Cluster. F.C. received also access to the NCSA \textit{Blue Waters} Cluster via the NSF AST--1516150 grant and to the TACC \texttt{Frontera} Cluster via the NSF PHI--1707946 grant. F.C. has been partially supported by the NASA TCAN 80NSSC18K1488 grant for a three--months visiting period at RIT. F. C. acknowledges also Dr. V. Mewes, Prof. Y. Zlochower, Prof. M. Campanelli and Prof. C. Lousto for interesting scientific discussions. J.\,V.\,K.~kindly acknowledges the CARIPARO Foundation (https://www.fondazionecariparo.it) for funding his PhD fellowship within the PhD School in Physics at the University of Padova.

\appendix

\section{Convergence Study}

\begin{figure}[t!]
	\centering
	\begin{minipage}{0.7\textwidth}
		\centering
		\includegraphics[width=1\linewidth]{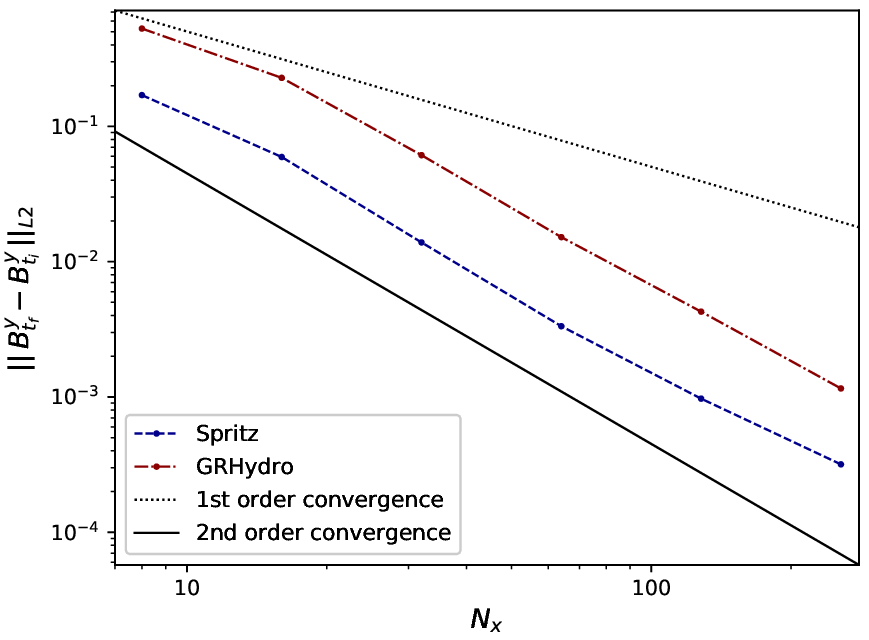} 
	\end{minipage}%

	\caption{Results of the Alfven Wave test simulation. The $L2$--norm of the error is plotted for both the \texttt{Spritz} code (blue--dashed line) and the \texttt{GRHydro} public code (red--dot--dashed line), along with reference curves for $1$st (black--dotted line) and $2$nd (black--solid line) order convergence.}
	\label{Fig19}
\end{figure}

In order to estimate the convergence order of our code, we decided to perform the so--called Alfv\'en Wave Test and compare the results with \texttt{GRHydro} (see \cite{mosta2013grhydro} for details). This test consists in the advection of a low--amplitude, circularly--polarized Alfv\'en wave across the domain. We used the same initial conditions of \cite{mosta2013grhydro}, namely: the wave amplitude $A_0 = 1.0$, the fluid rest--mass density $\rho = 1.0$, the fluid pressure $p_{\rm gas} = 1.0$, the $x$--component of velocity $v_x = 0.0$, the $x$--component of the magnetic field $B^x = 1.0$. We used a $\Gamma = \frac{5}{3}$ ideal fluid EOS and evolved for $1$ period. The convergence is studied considering several different resolutions along the $x$--axis, namely $N_x = 8$, $16$, $32$, $64$, $128$, and $256$, and using $x \in [-0.5, 2.5]$. The results for the $L2$--norm of the difference between the initial and final values of the $y$--component of the magnetic field, $B^y$, are presented in~\Fref{Fig19}. The initial wave is centered in $x=0$ and the final values of $B^y$ are compared with the initial profile. There, an overall $2$nd order convergence can be observed.

\section*{References}
\bibliographystyle{unsrt}
\bibliography{references}

\begin{thebibliography}{10}

\bibitem{anton2006numerical}
Ant{\'o}n L, Zanotti O, Miralles~J A, Mart{\'\i}~J M, Ib{\'a}{\~n}ez~J M,
  Font~J A, and Pons~J A.
\newblock Numerical 3+1 general relativistic magnetohydrodynamics: a local
  characteristic approach.
\newblock {\em ApJ}, 637(1):296, 2006.

\bibitem{giacomazzo2007whiskymhd}
Giacomazzo B and Rezzolla L.
\newblock Whiskymhd: a new numerical code for general relativistic
  magnetohydrodynamics.
\newblock {\em Class. Quantum Grav.}, 24(12):235, 2007.

\bibitem{mosta2013grhydro}
M{\"o}sta P, Mundim~B C, Faber~J A, Haas R, Noble~S C, Bode T, L{\"o}ffler F,
  Ott~C D, Reisswig C, and Schnetter E.
\newblock Grhydro: a new open-source general-relativistic magnetohydrodynamics
  code for the einstein toolkit.
\newblock {\em Class. Quantum Grav.}, 31(1):015005, 2013.

\bibitem{etienne2015illinoisgrmhd}
Etienne~Z B, Paschalidis V, Haas R, M{\"o}sta P, and Shapiro~S L.
\newblock Illinoisgrmhd: an open-source, user-friendly grmhd code for dynamical
  spacetimes.
\newblock {\em Class. Quantum Grav.}, 32(17):175009, 2015.

\bibitem{Liu2008}
Liu~Y T, Shapiro~S L, Etienne~Z B, and Taniguchi K.
\newblock General relativistic simulations of magnetized binary neutron star
  mergers.
\newblock {\em PRD}, 78(2):024012, 2008.

\bibitem{Anderson2008}
Anderson M, Hirschmann~E W, Lehner L, Liebling~S L, Motl~P M, Neilsen D,
  Palenzuela C, and Tohline~J E.
\newblock Magnetized neutron-star mergers and gravitational-wave signals.
\newblock {\em PRL}, 100(19):191101, 2008.

\bibitem{giacomazzo2011accurate}
Giacomazzo B, Rezzolla L, and Baiotti L.
\newblock Accurate evolutions of inspiralling and magnetized neutron stars:
  Equal-mass binaries.
\newblock {\em PRD}, 83(4):044014, 2011.

\bibitem{Kiuchi2018}
Kiuchi K, Kyutoku K, Sekiguchi Y, and Shibata M.
\newblock Global simulations of strongly magnetized remnant massive neutron
  stars formed in binary neutron star mergers.
\newblock {\em PRD}, 97(12):124039, 2018.

\bibitem{Ciolfi2019}
Ciolfi R, Kastaun W, Kalinani~J V, and Giacomazzo B.
\newblock First 100 ms of a long-lived magnetized neutron star formed in a
  binary neutron star merger.
\newblock {\em PRD}, 100(2):023005, 2019.

\bibitem{Palenzuela2010}
Palenzuela C, Lehner L, and Liebling~S L.
\newblock Dual jets from binary black holes.
\newblock {\em Science}, 329(5994):927, 2010.

\bibitem{Giacomazzo2012}
Giacomazzo B, Baker~J G, Miller~M C, Reynolds~C S, and van Meter J~R.
\newblock General relativistic simulations of magnetized plasmas around merging
  supermassive black holes.
\newblock {\em ApJ Lett.}, 752(1):15, 2012.

\bibitem{Farris2012}
Farris~B D, Gold R, Paschalidis V, Etienne~Z B, and Shapiro~S L.
\newblock Binary black-hole mergers in magnetized disks: Simulations in full
  general relativity.
\newblock {\em PRL}, 109(22):221102, 2012.

\bibitem{Gold2014}
Gold R, Paschalidis V, Ruiz M, Shapiro~S L, Etienne~Z B, and Pfeiffer~H P.
\newblock Accretion disks around binary black holes of unequal mass: General
  relativistic mhd simulations of postdecoupling and merger.
\newblock {\em PRD}, 90(10):104030, 2014.

\bibitem{Kelly2017}
Kelly~B J, Baker~J G, Etienne~Z B, Giacomazzo B, and Schnittman J.
\newblock Prompt electromagnetic transients from binary black hole mergers.
\newblock {\em PRD}, 96(12):123003, 2017.

\bibitem{dascoli2018}
d'Ascoli S, Noble~S C, Bowen~D B, Campanelli M, Krolik~J H, and Mewes V.
\newblock Electromagnetic emission from supermassive binary black holes
  approaching merger.
\newblock {\em ApJ}, 865(2):140, 2018.

\bibitem{EHTCodeComparison}
Chatterjee~K Porth~O et~al.
\newblock The event horizon general relativistic magnetohydrodynamic code
  comparison project.
\newblock {\em \APJ Suppl.}, 243(2):26, 2019.

\bibitem{Kawamura2016}
Kawamura T, Giacomazzo B, Kastaun W, Ciolfi R, Endrizzi A, Baiotti L, and Perna
  R.
\newblock Binary neutron star mergers and short gamma-ray bursts: Effects of
  magnetic field orientation, equation of state, and mass ratio.
\newblock {\em PRD}, 94:064012, 2016.

\bibitem{Ciolfi2017}
Ciolfi R, Kastaun W, Giacomazzo B, Endrizzi A, Siegel~D M, and Perna R.
\newblock General relativistic magnetohydrodynamic simulations of binary
  neutron star mergers forming a long-lived neutron star.
\newblock {\em PRD}, 95:063016, 2017.

\bibitem{Paschalidis2015}
Paschalidis V, Ruiz M, and Shapiro~S L.
\newblock Relativistic simulations of black hole-neutron star coalescence: The
  jet emerges.
\newblock {\em ApJ Lett.}, 806(1):14, 2015.

\bibitem{Ruiz2016}
Ruiz M, Lang~R N, Paschalidis V, and Shapiro~S L.
\newblock Binary neutron star mergers: A jet engine for short gamma-ray bursts.
\newblock {\em ApJ Lett.}, 824(1):6, 2016.

\bibitem{PhysRevLett.119.161101}
Abbott~B P, Abbott R, et~al.
\newblock Gw170817: Observation of gravitational waves from a binary neutron
  star inspiral.
\newblock {\em PRL}, 119:161101, 2017.

\bibitem{LVC-GRB}
Abbott~B P, Abbott R, et~al.
\newblock Gravitational waves and gamma-rays from a binary neutron star merger:
  Gw170817 and grb 170817a.
\newblock {\em ApJ Lett}, 848:13, 2017.

\bibitem{Schnittman2013}
Schnittman~J D.
\newblock Astrophysics of super-massive black hole mergers.
\newblock {\em Class. Quantum Grav.}, 30(24):244007, 2013.

\bibitem{Amaro-Seoane2017}
Amaro-Seoane P, Audley H, et~al.
\newblock Laser interferometer space antenna.
\newblock {\em arXiv:1702.00786}, 2017.

\bibitem{Giacomazzo2009}
Giacomazzo B, Rezzolla L, and Baiotti L.
\newblock Can magnetic fields be detected during the inspiral of binary neutron
  stars?
\newblock {\em MNRAS}, 399(1):164, 2009.

\bibitem{Rezzolla2011}
Rezzolla L, Giacomazzo B, Baiotti L, Granot J, Kouveliotou C, and Aloy~M A.
\newblock The missing link: Merging neutron stars naturally produce jet-like
  structures and can power short gamma-ray bursts.
\newblock {\em ApJ Lett.}, 732(1):6, 2011.

\bibitem{GiacomazzoPerna2013}
Giacomazzo B and Perna R.
\newblock Formation of stable magnetars from binary neutron star mergers.
\newblock {\em ApJ Lett.}, 771(2):26, 2013.

\bibitem{Giacomazzo2015}
Giacomazzo B, Zrake J, Duffell~P C, MacFadyen~A I, and Perna R.
\newblock Producing magnetar magnetic fields in the merger of binary neutron
  stars.
\newblock {\em ApJ}, 809(1):39, 2015.

\bibitem{Endrizzi2016}
Endrizzi A, Ciolfi R, Giacomazzo B, Kastaun W, and Kawamura T.
\newblock General relativistic magnetohydrodynamic simulations of binary
  neutron star mergers with the apr4 equation of state.
\newblock {\em Class. Quantum Grav.}, 33:164001, 2016.

\bibitem{Giacomazzo2012BBH}
B~{Giacomazzo}, J.~G. {Baker}, M.~C. {Miller}, C.~S. {Reynolds}, and J.~R. {van
  Meter}.
\newblock {General Relativistic Simulations of Magnetized Plasmas around
  Merging Supermassive Black Holes}.
\newblock {\em ApJ Letters}, 752(1):L15, Jun 2012.

\bibitem{Read2009}
Read~J S, Lackey~B D, Owen~B J, and Friedman~J L.
\newblock Constraints on a phenomenologically parametrized neutron-star
  equation of state.
\newblock {\em PRD}, 79(12):124032, 2009.

\bibitem{zenodo-spritz}
Cipolletta F., Kalinani~J. V., Giacomazzo B., and Ciolfi R.
\newblock The spritz code.
\newblock {\em Zenodo}, DOI:
  10.5281/zenodo.3689752:\url{http://dx.doi.org/10.5281/zenodo.3689752}, 2020.

\bibitem{baumgarte2010numerical}
Baumgarte~T W and Shapiro~S L.
\newblock {\em Numerical relativity: solving Einstein's equations on the
  computer}.
\newblock (Cambridge University Press), 2010.

\bibitem{banyuls1997numerical}
Banyuls F, Font~J A, Ib{\'a}{\~n}ez~J M, Mart{\'\i}~J M, and Miralles~J A.
\newblock Numerical $\{$3+1$\}$ general relativistic hydrodynamics: A local
  characteristic approach.
\newblock {\em ApJ}, 476(1):221, 1997.

\bibitem{marti1991numerical}
Mart{\'\i}~J M, Ib{\'a}nez~J M, and Miralles~J A.
\newblock Numerical relativistic hydrodynamics: local characteristic approach.
\newblock {\em PRD}, 43(12):3794, 1991.

\bibitem{baiotti2005three}
Baiotti L, Hawke I, Montero~P J, L{\"o}ffler F, Rezzolla L, Stergioulas N,
  Font~J A, and Seidel E.
\newblock Three-dimensional relativistic simulations of rotating neutron-star
  collapse to a kerr black hole.
\newblock {\em PRD}, 71(2):024035, 2005.

\bibitem{BaiottiPhDThesis}
Baiotti L.
\newblock {\em Numerical relativity simulations of non-vacuum spacetimes in
  three dimensions}.
\newblock PhD thesis, SISSA, 2004.
\newblock http://hdl.handle.net/20.500.11767/3994.

\bibitem{feynman1979feynman}
Feynman~R P, Leighton~R B, and Sands M.
\newblock {\em The Feynman lectures on physics, vol. 2: Mainly electromagnetism
  and matter}.
\newblock (Addison-Wesley), 1979.

\bibitem{baumgarte2003collapse}
Baumgarte~T W and Shapiro~S L.
\newblock Collapse of a magnetized star to a black hole.
\newblock {\em ApJ}, 585(2):930, 2003.

\bibitem{baumgarte2003general}
Baumgarte~T W and Shapiro~S L.
\newblock General relativistic magnetohydrodynamics for the numerical
  construction of dynamical spacetimes.
\newblock {\em ApJ}, 585(2):921, 2003.

\bibitem{etienne2012relativistic}
Etienne~Z B, Paschalidis V, Liu~Y T, and Shapiro~S L.
\newblock Relativistic magnetohydrodynamics in dynamical spacetimes: Improved
  electromagnetic gauge condition for adaptive mesh refinement grids.
\newblock {\em PRD}, 85(2):024013, 2012.

\bibitem{etienne2010relativistic}
Etienne~Z B, Liu~Y T, and Shapiro~S L.
\newblock Relativistic magnetohydrodynamics in dynamical spacetimes: A new
  adaptive mesh refinement implementation.
\newblock {\em PRD}, 82(8):084031, 2010.

\bibitem{ETKpaper}
L{\"o}ffler F, Faber J, Bentivegna E, Bode T, Diener P, Haas R, Hinder I,
  Mundim~B C, Ott~C D, Schnetter E, Allen G, Campanelli M, and Laguna P.
\newblock The einstein toolkit: a community computational infrastructure for
  relativistic astrophysics.
\newblock {\em Class. Quantum Grav.}, 29(11):115001, 2012.

\bibitem{EinsteinToolkit:2019_10}
Babiuc-Hamilton M, Brandt~S R, et~al.
\newblock The einstein toolkit, 2019.
\newblock To find out more, visit http://einsteintoolkit.org.

\bibitem{harten1983upstream}
Harten A, Lax~P D, and van Leer~B.
\newblock On upstream differencing and godunov-type schemes for hyperbolic
  conservation laws.
\newblock {\em SIAM Rev.}, 25(1):35, 1983.

\bibitem{toro2013riemann}
Toro~E F.
\newblock {\em Riemann solvers and numerical methods for fluid dynamics: a
  practical introduction}.
\newblock (Springer Science \& Business Media), 2013.

\bibitem{del2003efficient}
Del~Zanna L, Bucciantini N, and Londrillo P.
\newblock An efficient shock-capturing central-type scheme for multidimensional
  relativistic flows-ii. magnetohydrodynamics.
\newblock {\em A\&A}, 400(2):397, 2003.

\bibitem{colella1984piecewise}
Colella P and Woodward~P R.
\newblock The piecewise parabolic method (ppm) for gas-dynamical simulations.
\newblock {\em J. Comput. Phys}, 54(1):174, 1984.

\bibitem{balsara1999staggered}
Balsara~D S and Spicer~D S.
\newblock A staggered mesh algorithm using high order godunov fluxes to ensure
  solenoidal magnetic fields in magnetohydrodynamic simulations.
\newblock {\em J. Comput. Phys}, 149(2):270, 1999.

\bibitem{evans88}
Evans~C R and Hawley~J F.
\newblock Simulation of magnetohydrodynamic flows: A constrained transport
  method.
\newblock {\em ApJ}, 332:659, 1988.

\bibitem{Balsara2001}
Balsara~D S.
\newblock Divergence-free adaptive mesh refinement for magnetohydrodynamics.
\newblock {\em J. Comput. Phys}, 174(2):614, 2001.

\bibitem{balsara2001total}
Balsara D.
\newblock Total variation diminishing scheme for relativistic
  magnetohydrodynamics.
\newblock {\em ApJS}, 132(1):83, 2001.

\bibitem{Pollney2011}
Denis {Pollney}, Christian {Reisswig}, Erik {Schnetter}, Nils {Dorband}, and
  Peter {Diener}.
\newblock {High accuracy binary black hole simulations with an extended wave
  zone}.
\newblock {\em Phys. Rev. D}, 83(4):044045, Feb 2011.

\bibitem{Reisswig2013}
C.~{Reisswig}, R.~{Haas}, C.~D. {Ott}, E.~{Abdikamalov}, P.~{M{\"o}sta},
  D.~{Pollney}, and E.~{Schnetter}.
\newblock {Three-dimensional general-relativistic hydrodynamic simulations of
  binary neutron star coalescence and stellar collapse with multipatch grids}.
\newblock {\em Phys. Rev. D}, 87(6):064023, Mar 2013.

\bibitem{noble2006primitive}
Noble~S C, Gammie~C F, McKinney~J C, and Del~Zanna L.
\newblock Primitive variable solvers for conservative general relativistic
  magnetohydrodynamics.
\newblock {\em ApJ}, 641(1):626, 2006.

\bibitem{Siegel2018}
Siegel~D M, M{\"o}sta P, Desai D, and Wu~S.
\newblock Recovery schemes for primitive variables in general-relativistic
  magnetohydrodynamics.
\newblock {\em ApJ}, 859(1):71, 2018.

\bibitem{horedt2004polytropes}
Horedt~G P.
\newblock {\em Polytropes: applications in astrophysics and related fields}.
\newblock (Springer Science \& Business Media), 2004.

\bibitem{COMPOSE}
Compose, https://compose.obspm.fr/.

\bibitem{cipolletta2015fast}
Cipolletta F, Cherubini C, Filippi S, Rueda~J A, and Ruffini R.
\newblock Fast rotating neutron stars with realistic nuclear matter equation of
  state.
\newblock {\em PRD}, 92(2):023007, 2015.

\bibitem{Carpet}
Carpet, https://carpetcode.org/.

\bibitem{schnetter2004evolutions}
Schnetter E, Hawley~S H, and Hawke I.
\newblock Evolutions in 3d numerical relativity using fixed mesh refinement.
\newblock {\em Class. Quantum Grav.}, 21(6):1465, 2004.

\bibitem{Brown:2008sb}
J.~D. Brown, P.~Diener, O.~Sarbach, E.~Schnetter, and M.~Tiglio.
\newblock {Turduckening black holes: an analytical and computational study}.
\newblock {\em Phys. Rev. D}, 79:044023, 2009.

\bibitem{Kranc:web}
{Kranc}: {Kranc} assembles numerical code.
\newblock http://kranccode.org/.

\bibitem{McLachlan:web}
{McLachlan}, a public {BSSN} code.
\newblock http://www.cct.lsu.edu/~eschnett/McLachlan/.

\bibitem{baumgarte1998numerical}
Baumgarte~T W and Shapiro~S L.
\newblock Numerical integration of einstein’s field equations.
\newblock {\em PRD}, 59(2):024007, 1998.

\bibitem{nakamura1987general}
Nakamura T, Oohara K, and Kojima Y.
\newblock General relativistic collapse to black holes and gravitational waves
  from black holes.
\newblock {\em PROG THEOR PHYS SUPP}, 90:1, 1987.

\bibitem{shibata1995evolution}
Shibata M and Nakamura T.
\newblock Evolution of three-dimensional gravitational waves: Harmonic slicing
  case.
\newblock {\em PRD}, 52(10):5428, 1995.

\bibitem{alcubierre2003gauge}
Alcubierre M, Br{\"u}gmann B, Diener P, Koppitz M, Pollney D, Seidel E, and
  Takahashi R.
\newblock Gauge conditions for long-term numerical black hole evolutions
  without excision.
\newblock {\em PRD}, 67(8):084023, 2003.

\bibitem{alcubierre2000towards}
Alcubierre M, Br{\"u}gmann B, Dramlitsch T, Font~J A, Papadopoulos P, Seidel E,
  Stergioulas N, and Takahashi R.
\newblock Towards a stable numerical evolution of strongly gravitating systems
  in general relativity: The conformal treatments.
\newblock {\em PRD}, 62(4):044034, 2000.

\bibitem{beckwith2011second}
Beckwith K and Stone~J M.
\newblock A second-order godunov method for multi-dimensional relativistic
  magnetohydrodynamics.
\newblock {\em ApJS}, 193(1):6, 2011.

\bibitem{giacomazzo2006exact}
Giacomazzo B and Rezzolla L.
\newblock The exact solution of the riemann problem in relativistic
  magnetohydrodynamics.
\newblock {\em J. Fluid Mech.}, 562:223, 2006.

\bibitem{komissarov1999godunov}
Komissarov~S S.
\newblock A godunov-type scheme for relativistic magnetohydrodynamics.
\newblock {\em MNRAS}, 303(2):343, 1999.

\bibitem{del2007echo}
Del~Zanna L, Zanotti O, Bucciantini N, and Londrillo P.
\newblock Echo: a eulerian conservative high-order scheme for general
  relativistic magnetohydrodynamics and magnetodynamics.
\newblock {\em A\&A}, 473(1):11, 2007.

\bibitem{toth2000b}
T{\'o}th G.
\newblock The $\nabla$ {\textperiodcentered} b = 0 constraint in
  shock-capturing magnetohydrodynamics codes.
\newblock {\em J. Comput. Phys}, 161(2):605, 2000.

\bibitem{devore1991flux}
DeVore~C R.
\newblock Flux-corrected transport techniques for multidimensional compressible
  magnetohydrodynamics.
\newblock {\em J. Comput. Phys}, 92(1):142, 1991.

\bibitem{gardiner2005unsplit}
Gardiner~T A and Stone~J M.
\newblock An unsplit godunov method for ideal mhd via constrained transport.
\newblock {\em J. Comput. Phys}, 205(2):509, 2005.

\bibitem{stone2008athena}
Stone~J M, Gardiner~T A, Teuben P, Hawley~J F, and Simon~J B.
\newblock Athena: a new code for astrophysical mhd.
\newblock {\em ApJS}, 178(1):137, 2008.

\bibitem{cerda2008new}
Cerd{\'a}-Dur{\'a}n P, Font~J A, Ant{\'o}n L, and M{\"u}ller E.
\newblock A new general relativistic magnetohydrodynamics code for dynamical
  spacetimes.
\newblock {\em A\&A}, 492(3):937, 2008.

\bibitem{cerda2007general}
Cerd{\'a}-Dur{\'a}n P, Font~J A, and Dimmelmeier H.
\newblock General relativistic simulations of passive-magneto-rotational core
  collapse with microphysics.
\newblock {\em A\&A}, 474(1):169, 2007.

\bibitem{Mewes2020}
Vassilios {Mewes}, Yosef {Zlochower}, Manuela {Campanelli}, Thomas~W.
  {Baumgarte}, Zachariah~B. {Etienne}, Federico~G. {Lopez Armengol}, and
  Federico {Cipolletta}.
\newblock {Numerical relativity in spherical coordinates: A new dynamical
  spacetime and general relativistic MHD evolution framework for the Einstein
  Toolkit}.
\newblock {\em Phys. Rev. D}, 101:104007, May 2020.

\bibitem{oppenheimer1939massive}
Oppenheimer~J R and Volkoff~G M.
\newblock On massive neutron cores.
\newblock {\em Phys. Rev.}, 55(4):374, 1939.

\bibitem{tolman1939static}
Tolman~R C.
\newblock Static solutions of einstein's field equations for spheres of fluid.
\newblock {\em Phys. Rev.}, 55(4):364, 1939.

\bibitem{cowling1941non}
Cowling~T G.
\newblock The non-radial oscillations of polytropic stars.
\newblock {\em MNRAS}, 101:367, 1941.

\bibitem{lindblom1990accuracy}
Lindblom L and Splinter~R J.
\newblock The accuracy of the relativistic cowling approximation.
\newblock {\em ApJ}, 348:198, 1990.

\bibitem{1969ApJ...158..997T}
Thorne~K S.
\newblock Nonradial pulsation of general-relativistic stellar models.iv. the
  weakfield limit.
\newblock {\em ApJ}, 158:997, 1969.

\bibitem{font2000non}
Font~J A, Stergioulas N, and Kokkotas~K D.
\newblock Non-linear hydrodynamical evolution of rotating relativistic stars:
  numerical methods and code tests.
\newblock {\em MNRAS}, 313(4):678, 2000.

\bibitem{font2002three}
Font~J A, Goodale T, Iyer S, Miller M, Rezzolla L, Seidel E, Stergioulas N,
  Suen W, and Tobias M.
\newblock Three-dimensional numerical general relativistic hydrodynamics. ii.
  long-term dynamics of single relativistic stars.
\newblock {\em PRD}, 65(8):084024, 2002.

\bibitem{most2019beyond}
E.~R. {Most}, L.~Jens {Papenfort}, and L.~{Rezzolla}.
\newblock {Beyond second-order convergence in simulations of magnetized binary
  neutron stars with realistic microphysics}.
\newblock {\em MNRAS}, 490(3):3588--3600, December 2019.

\bibitem{palenzuela2015effects}
Palenzuela C, Liebling~S L, Neilsen D, Lehner L, Caballero~O L, O'Connor E, and
  Anderson M.
\newblock Effects of the microphysical equation of state in the mergers of
  magnetized neutron stars with neutrino cooling.
\newblock {\em PRD}, 92(4):044045, 2015.

\end{thebibliography}

\end{document}